\documentclass[groupedaddress,superscriptaddress,aps,prd,preprintnumbers,onecolumn,10pt,fleqn,nofootinbib,nobibnote,eqsecnum,floatfix,showpacs,preprintnumbers]{revtex4-2}

\usepackage{epsfig}
\usepackage{graphicx}
\usepackage{palatino}
\usepackage[english]{babel}
\usepackage{hyphenat}
\usepackage{amsmath}
\usepackage{amssymb}
\usepackage{mathtools}
\usepackage{mathrsfs}
\usepackage{bbm} 
\usepackage{bm}
\usepackage{slashed}
\usepackage{epstopdf}
\usepackage{xcolor}
\usepackage{booktabs}
\usepackage{float}
\definecolor{lcolor}{rgb}{0.,0.0,0.}
\definecolor{citcolor}{rgb}{0,0.,0.5}
\usepackage[breaklinks,colorlinks,urlcolor=blue,citecolor=blue,linkcolor=blue]{hyperref}
\usepackage{multirow}
\usepackage{ltablex}
\usepackage{soul}

\newcommand{\nlo}{{\rm \scriptscriptstyle NLO}}

\newcommand{\lo}{{\rm \scriptscriptstyle LO}}

\newcommand{\eqn}[1]{Eq.\,\eqref{#1}}
\newcommand{\beq}{\begin{equation}}
\newcommand{\eeq}{\end{equation}}
\newcommand{\bal}{\begin{align}}
\newcommand{\eal}{\end{align}}

\newcommand{\minus}{\!-\!}

\long\def\comment#1{ }

\newcommand{\del}{\partial}

\newcommand{\order}[1]{\mathcal{O}{(#1)}}
\newcommand{\abar}{\bar{\alpha}_s}
\newcommand{\abarz}{\bar{\alpha}_0}
\newcommand{\nn}{\nonumber\\}
\newcommand{\rmd}{{\rm d}}
\newcommand{\dif}{{\rm d}}
\newcommand{\rme}{{\rm e}}
\newcommand{\bk}{\bm{k}}

\newcommand{\bx}{\bm{x}}
\newcommand{\by}{\bm{y}}

\newcommand{\bz}{\bm{z}}

\newcommand{\br}{\bm{r}}
\newcommand{\bb}{\bm{b}}
\newcommand{\bbp}{\bm{b}_\perp}
\newcommand{\bbb}{\overline{\bm{b}}}
\newcommand{\bt}{b_\perp}

\newcommand{\mcal}{\mathcal}
\newcommand{\bK}{\bm{K}_\perp}
\newcommand{\bP}{\bm{P}_\perp}
\newcommand{\bR}{\bm{R}}
\newcommand{\KT}{K_\perp}
\newcommand{\PT}{P_\perp}
\newcommand{\kg}{k_{g\perp}}
\newcommand{\bkg}{\bm{k_{g}}}
\newcommand{\lt}{\ell_{\perp}}
\newcommand{\ellt}{\bm{\ell}_\perp}

\newcommand{\tF}{\tilde{\mcal{F}}_g}

\begin{document}

\title{Evolution of the transverse-momentum dependent  gluon distribution at small $x$}
\author{Paul Caucal}
\email{caucal@subatech.in2p3.fr}
 \affiliation{SUBATECH UMR 6457 (IMT Atlantique, Université de Nantes, IN2P3/CNRS), 4 rue Alfred Kastler, 44307 Nantes, France}
\author{Edmond Iancu}
\email{edmond.iancu@ipht.fr}
 \affiliation{Université Paris-Saclay, CNRS, CEA, Institut de physique théorique, F-91191, Gif-sur-Yvette, France}

\begin{abstract}
 Using the  colour dipole picture for photon-nucleus interactions at small $x$ together with the 
Color Glass Condensate (CGC) effective theory,
 we demonstrate that the next-to-leading (NLO) order corrections to the cross-section for the inclusive production of a pair of hard jets encode not only the JIMWLK evolution with decreasing $x$, but also the DGLAP evolution of the gluon 
 distribution function and the CSS evolution of the gluon transverse momentum dependent (TMD) distribution.
 The emergent CSS equation takes the form of a rate equation describing the evolution of the dijet 
 distribution in the transverse momentum imbalance $K_\perp$ when increasing the dijet  relative momentum $P_\perp$.
  All three types of evolution become important when both $P_\perp$ and $K_\perp$
 are much larger than the nuclear saturation momentum $Q_s(x)$ and we propose a framework which 
 encompasses all of them. 
  The solution to the JIMWLK equation provides the source term for the DGLAP
 evolution with increasing $K_\perp$, which in turn generates the initial condition for the CSS evolution
 with increasing $P_\perp$.

\end{abstract}

\maketitle

\clearpage
\tableofcontents
\vspace{10 mm}

\section{Introduction}
One of the main challenges in QCD at high energy consists in building a theoretical framework allowing for a simultaneous resummation of  small-$x$, DGLAP and Sudakov 
logarithms in initial-state parton distribution functions (PDF)~\cite{Mueller:2012uf,Mueller:2013wwa,Balitsky:2015qba,Balitsky:2016dgz,Zhou:2016tfe,Xiao:2017yya,Hentschinski:2021lsh,Boussarie:2021wkn,Fu:2023jqv,Mukherjee:2023snp}. In this paper, through the example of inclusive back-to-back dijet production in photon-nucleus ($\gamma A$)  interactions at high energy,  we would like to argue that TMD factorisation \cite{Collins:2011zzd,Boussarie:2023izj} at small Bjorken $x$
is an appropriate framework to achieve this goal. Besides being interesting for the phenomenology (notably for
 ultra-peripheral Pb+Pb collisions at the LHC  \cite{ATLAS:2022cbd,CMS:2022lbi,ATLAS:2024mvt} and electron-nucleus ($eA$) deep inelastic scattering at the EIC~\cite{Accardi:2012qut,Aschenauer:2017jsk,AbdulKhalek:2022hcn}), this dijet process is also well suited for our purposes: it offers the right combination of longitudinal and transverse momentum scales to create the premises for both the high-energy and the TMD factorisations.


The high-energy factorisation for $\gamma A$ interactions relies on the colour dipole picture (CDP)~\cite{Kopeliovich:1981pz,Bertsch:1981py,Mueller:1989st,Nikolaev:1990ja}, modernly embedded in 
the CGC effective theory~\cite{Iancu:2002xk,Iancu:2003xm,Gelis:2010nm,Kovchegov:2012mbw} and the associated  BK/JIMWLK evolution with  decreasing $x$~\cite{Balitsky:1995ub,Kovchegov:1999yj,JalilianMarian:1997jx,JalilianMarian:1997gr,Kovner:2000pt,Weigert:2000gi,Iancu:2000hn,Iancu:2001ad,Ferreiro:2001qy}. Both the cross-section for dijet production in the CDP~\cite{Caucal:2021ent,Taels:2022tza,Bergabo:2022tcu,Iancu:2022gpw} and the BK/JIMWLK evolution~\cite{Balitsky:2008zza,Balitsky:2013fea,Kovner:2013ona,Kovner:2014lca,Lublinsky:2016meo,Lappi:2020srm,Dai:2022imf} are presently known to next-to-leading order (NLO).  TMD factorisation at small $x$
 has been demonstrated to emerge from the CGC effective theory for a variety of multi-scale processes, including the dijet process relevant for us here~\cite{Marquet:2009ca,Dominguez:2010xd,Dominguez:2011wm,Metz:2011wb,Dominguez:2011br,Dumitru:2015gaa,Kotko:2015ura,Marquet:2016cgx,vanHameren:2016ftb,Marquet:2017xwy,Albacete:2018ruq,Dumitru:2018kuw,Boussarie:2021ybe,Kotko:2017oxg,Roy:2019hwr,Roy:2019cux,Iancu:2021rup,Hatta:2022lzj,Iancu:2022lcw,Hauksson:2024bvv}.
At   leading order (LO), the TMD factorisation for dijet production at small $x$ \cite{Dominguez:2010xd,Dominguez:2011wm} involves the Weiszäcker-Williams (WW) gluon TMD, which physically represents the unintegrated gluon distribution of the target~\cite{McLerran:1993ka,McLerran:1993ni,McLerran:1998nk}. Recent studies of the NLO  corrections within the CGC effective theory \cite{Taels:2022tza,Caucal:2022ulg,Caucal:2023nci,Caucal:2023fsf} have revealed 
the high-energy evolution of the WW gluon TMD \cite{Dominguez:2011gc} together with the  Sudakov effects associated with final state gluon emissions~\cite{Mueller:2012uf,Mueller:2013wwa}.

 In this paper we will revisit the NLO  corrections to dijet production with the purpose of demonstrating that they encode two additional types of quantum evolution,  which reflect the hierarchy of  transverse momentum scales in the problem: the Dokshitzer-Gribov-Lipatov-Altarelli-Parisi (DGLAP) evolution of the gluon PDF~\cite{Gribov:1972ri,Altarelli:1977zs,Dokshitzer:1977sg}  and  the Collins-Soper-Sterman (CSS) evolution of the gluon TMD~\cite{Collins:1981uk,Collins:1981uw,Collins:1984kg,Collins:2011zzd}. These evolutions are well known in the context of the collinear/TMD factorisations at moderate $x$ \cite{Collins:2011zzd,Boussarie:2023izj}, but to our knowledge it is for the first time that they are unveiled within the CGC approach at small $x$.
  As we shall see, the DGLAP and CSS equations acquire special features  which are specific to the  small-$x$ context at hand: they appear on top of the BK/JIMWLK evolution, from which they inherit source terms and/or boundary conditions.

To exhibit these new evolutions, we shall perform an alternative calculation of  the NLO corrections, which includes a leading-twist approximation that exploits the separation of transverse momentum scales in the problem. In general, the problem of dijet production in dilute-dense (here, photon-nucleus, of $\gamma^*A$) collisions is characterised by three important transverse momentum scales: the relative momentum $\bP$ of the two produced jets, their momentum imbalance $\bK$, and the target saturation momentum $Q_s(A,x)$, where $A$ is the nucleus mass number and $x$ is the longitudinal momentum fraction of the gluons from the target which are involved in the scattering. The jets are hard and back-to-back in the transverse plane where $P_\perp$ is much larger than both $K_\perp$ and $Q_s$. For this particular configuration and to leading order (LO) in perturbation theory, the CGC calculation demonstrates TMD factorisation \cite{Dominguez:2010xd,Dominguez:2011wm}: the cross-section for dijet production can be written as the product between a hard factor encoding the dependence upon $\bP$ and the Weiszäcker-Williams (WW) gluon TMD, which counts the number of gluons from the target with transverse momentum $\bK$ and longitudinal momentum fraction $x$. This factorisation holds up to kinematical corrections suppressed by powers of $K_\perp/P_\perp$ or of $Q_s/P_\perp$. The most interesting regime for studies of gluon saturation is when the momentum imbalance is {\it semi-hard}, i.e. comparable to the saturation scale, $K_\perp\sim Q_s(A,x)$. In that case the WW gluon TMD is sensitive to saturation effects that could be measured e.g. via dijet correlations in azimuthal angle. The situation is however complicated by higher-order corrections which are numerically large due to the large transverse phase-space, between  $Q_s$ and $P_\perp$, that is available to additional radiation.

\begin{figure}
   \hspace*{-0.4cm}  \centering
    \includegraphics[width=0.28\linewidth]{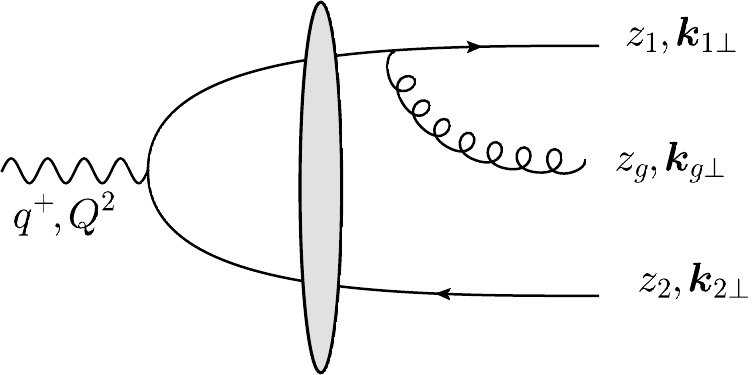}\quad
    \includegraphics[width=0.28\linewidth]{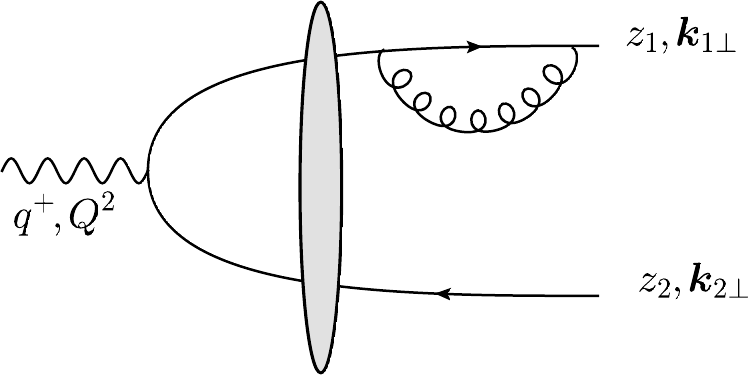}\quad
    \includegraphics[width=0.28\linewidth]{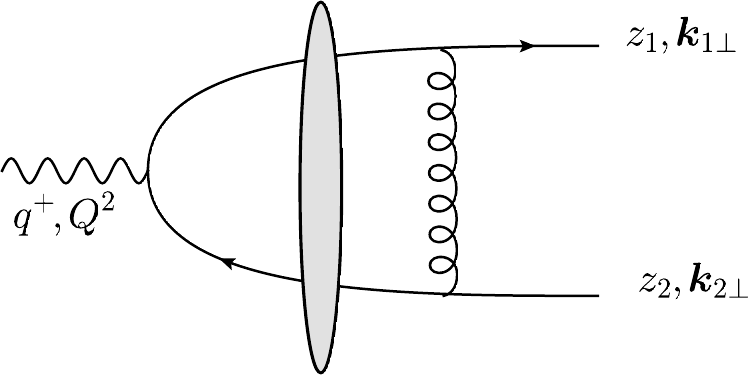}
    \caption{Feynman graphs contributing to quark-antiquark dijet production in $\gamma^*A$ collisions at NLO: one real gluon emissions and two virtual ones, all occurring in the final state (after the collision with the nucleus depicted as a shockwave).
    }
    \label{fig:gluon_after}
\end{figure}

The CGC calculation is based on the colour dipole picture in which the dijet formation (and, more generally, the radiative corrections) are computed as the quantum evolution of the {\it projectile}: long before the collision,  the virtual photon fluctuates into a quark-antiquark pair in a colour singlet state (a colour dipole) which is later put on-shell by its scattering off the dense gluon system in the target (represented by a Lorentz-contracted, random, colour background field --- a ``shockwave''). In this picture, the dominant NLO corrections to back-to-back dijet production come from gluon emissions by the quark or the antiquark in the final state (after the scattering with the shockwave), which have a double-logarithmic phase-space (in both transverse and longitudinal momenta); see Fig.~\ref{fig:gluon_after} for some examples of Feynman graphs.
The net result at NLO, after partial compensations between real and virtual gluon emissions, is negative and of order $\alpha_s\ln^2(P_\perp^2/K_\perp^2)$ --- the Sudakov double logarithm \cite{Mueller:2012uf,Mueller:2013wwa}. One expects this NLO result to exponentiate after resumming similar radiative corrections to all orders. Physically, this implies  a strong suppression of dijet production with $P_\perp\gg K_\perp$. Conversely, it favours dijet configurations where the momentum imbalance $K_\perp$ is closer to $P_\perp$ (and much larger than $Q_s$), thus reducing the influence of saturation \cite{Zheng:2014vka}. But in order for this effect to be firmly established, one needs a reliable calculation of the ``large'' radiative corrections --- those enhanced by powers of the transverse logarithm $\ln(P_\perp^2/K_\perp^2)$.

\begin{figure}
   \hspace*{-0.4cm}  \centering
    \includegraphics[width=0.28\linewidth]{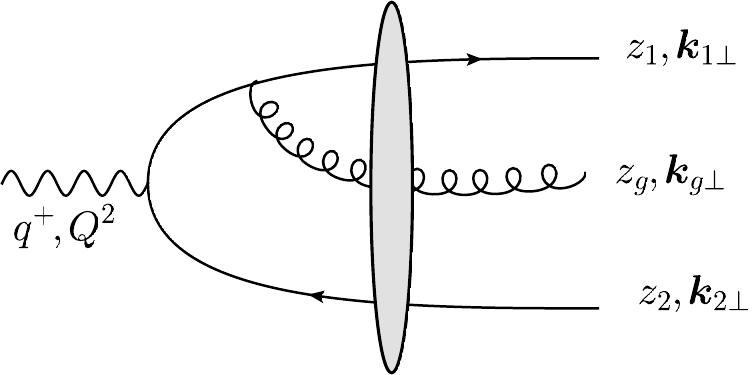}\quad
    \includegraphics[width=0.28\linewidth]{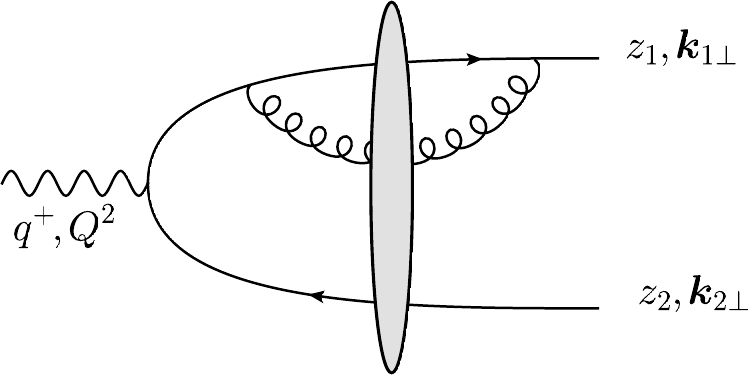}\quad
    \includegraphics[width=0.28\linewidth]{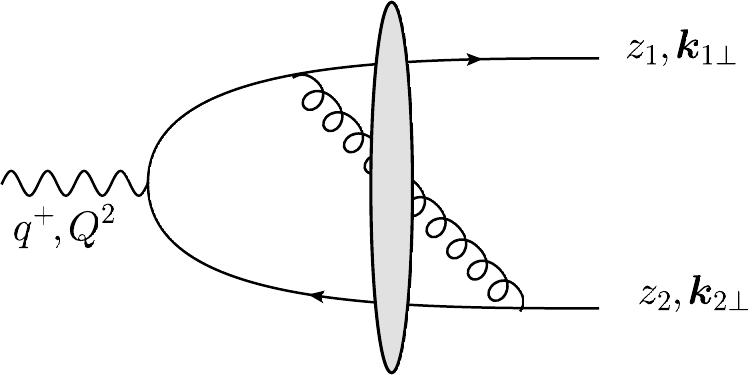}
    \caption{Similar to Fig.~\ref{fig:gluon_after}, except that the gluon emissions are now initiated before
    the collision with the nuclear target.}
    \label{fig:gluon_before}
\end{figure}

Besides the Sudakov double logarithm, one expects single logarithmic NLO corrections $\sim \alpha_s\ln(P_\perp^2/Q_s^2)$ associated with the DGLAP evolution of the gluon distribution in the target. Indeed, if the dijet imbalance is not measured, i.e. if the dijet cross-section is integrated over $K_\perp$ up to the hard scale $P_\perp$, the LO CGC result exhibits {\it collinear factorisation}: the integral of the gluon TMD provides the gluon PDF $xG(x, P_\perp^2)$, which surely must obey DGLAP evolution with increasing $P_\perp^2$. This evolution, that had not been identified so far in the CGC calculations (with the noticeable exception of diffractive single inclusive jet production in $\gamma^* A$ collisions \cite{Hauksson:2024bvv}), is naturally visualised as a change in the  distribution of the target gluons involved in the scattering ---  this a {\it target} evolution. So, at this level, one faces two important conceptual questions: \texttt{(i)} is the DGLAP evolution of the {\it target} gluon distribution encoded in the CGC approach (and if so, how to extract it ?), and  \texttt{(ii)} what is the relation between the Sudakov double logarithm (which in the CGC calculation is naturally associated with final-state radiation by the projectile) and the DGLAP evolution of the target ? That such a relation should exist, is strongly suggested by the experience with TMD factorisation at moderate values of $x$, where the Sudakov logarithms appear via end-point singularities  in the rapidity integrals over DGLAP splitting functions and they can be resummed to all orders by solving the CSS equation \cite{Collins:2011zzd,Boussarie:2023izj}.

It is our main purpose in this paper to answer the two questions above. Namely, we will demonstrate the emergence of the DGLAP and the CSS evolutions for the gluon distribution of the target via NLO calculations in the CGC effective theory. To that aim, we will have to consider both initial state and final state emissions, cf. Fig.~\ref{fig:gluon_after} and Fig.~\ref{fig:gluon_before}.
 We will explicitly compute only the ``real'' NLO corrections (those associated with the emission of a gluon in the final state), which have not been  computed in the approximations of interest. The corresponding virtual terms will be taken from the literature~\cite{Taels:2022tza,Caucal:2021ent,Caucal:2023nci,Caucal:2023fsf,Zhou:2018lfq}. To unveil the collinear (DGLAP and CSS) evolutions, we will consider a gluon  emission with relatively large transverse momentum $\kg$, within the range $\PT\gg \kg\gg Q_s$, whose recoil controls the dijet imbalance: $\KT\simeq\kg$. To contribute to the dijet cross-section, this emission will be further constrained to propagate at {\it large angles}, outside the cones defining the produced quark-antiquark jets. As we shall see, this condition introduces an upper limit on the longitudinal momentum fraction $z_g$ of the gluon w.r.t. the photon: $z_g\lesssim \KT/\PT$ (which in particular implies  $z_g\ll 1$). The collinear evolutions will be recognised after systematically expanding the NLO corrections to leading order in the small ratios $\KT/\PT$ and $Q_s/\KT$.

\begin{figure}[t]
    \centering
    \includegraphics[width=0.48\textwidth]{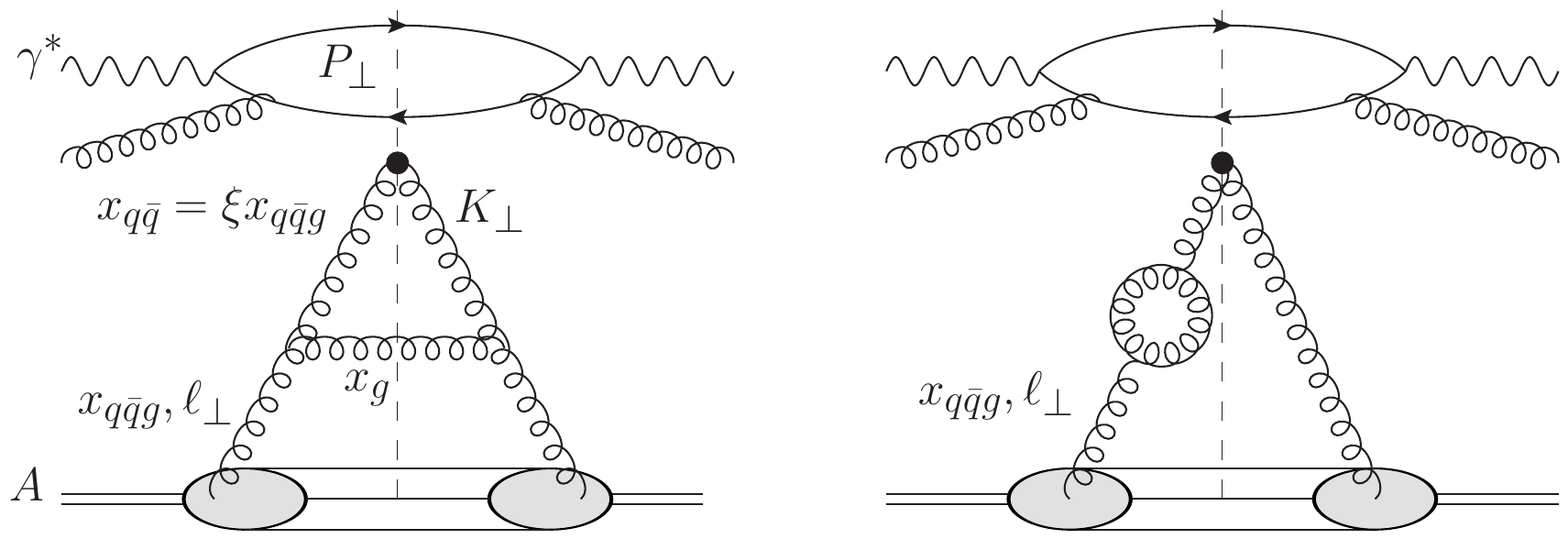}
    \caption{TMD factorisation for the dijet cross-section  at next-to-leading order in the target picture.}
    \label{fig:diagrams}
\end{figure}

What renders our construction particularly subtle is the fact that the NLO corrections are explicitly computed as gluon emissions by the {\it projectile} (the quark-antiquark fluctuation of the virtual photon), whereas the gluon TMD and its evolution refer to the wavefunction of the {\it nuclear target}.  
The longitudinal phase-space available to gluon emissions is differently parametrised in these two pictures 
--- projectile evolution and target evolution, respectively. This complicates the separation between the various types of evolution, that must be first performed in the projectile picture. To achieve this separation, it is crucial to observe~\cite{Caucal:2022ulg,Taels:2022tza,Caucal:2023fsf} that the correct high-energy evolution in the kinematics at hand is described by the {\it collinearly improved} version of the BK/JIMWLK equations \cite{Beuf:2014uia,Iancu:2015vea,Hatta:2016ujq,Ducloue:2019ezk},  in which successive gluon emissions by the projectile obey {\it time ordering}: their formation times are shorter and shorter \cite{Iancu:2015vea}. This condition restricts the longitudinal phase-space for the high energy evolution --- as we shall see, this is limited to {\it very} small longitudinal fractions $z_g\ll \KT^2/\PT^2$ --- and thus opens the phase-space for the DGLAP and CSS evolutions. In order for the latter to become manifest, one also needs to perform a change of longitudinal variables, from $z_g$ to $x_g$ (the gluon  longitudinal momentum fraction w.r.t.  nuclear target). A similar strategy  has been used in Ref.~\cite{Hauksson:2024bvv} in the context of diffractive jet production. The ultimate reason why this strategy works in the problem at hand is because the gluon emission is {\it soft} both w.r.t. the projectile and w.r.t. the target: $z_g\ll 1$ and  $x_g\ll 1$. Such soft gluons are emitted at mid-rapidities and can be transferred from the wavefunction of the projectile to that of the target. 

By following this strategy, we will demonstrate that TMD factorisation for back-to-back dijet production in $\gamma^*A$ collisions is preserved by the NLO corrections, as computed in the CGC effective theory and at  leading twist. Moreover, these corrections will be recognised as one step in the DGLAP evolution of the gluon PDF together with one step in the CSS evolution of the gluon TMD. 
In particular, we will reconstruct the DGLAP splitting function $P_{gg}(\xi)$ describing gluon splittings in the {\it target} and  for {\it generic} values for  $\xi$
from {\it soft} gluon emissions by the quark or the antiquark from the {\it projectile}.  We will more precisely identify a ``diagonal'' version of the CSS equations in which the two resolution scales (transverse and longitudinal) that are generally associated with the quantum evolution of a TMD are both controlled by a single, hard, transverse momentum: the dijet relative momentum $P_\perp$. This diagonal version naturally emerges from the CGC calculation, 
 where the hard scale $P_\perp$ effectively acts an ultraviolet cutoff on the virtual emissions and also controls the upper limit on the gluon longitudinal fraction $z_g$.  It has the additional virtue (only recently recognised within the TMD literature \cite{Ebert:2022cku,delRio:2024vvq}) to be consistent with the DGLAP evolution of the gluon PDF $xG(x, \PT^2)$, with the latter defined as the momentum integral of  the gluon TMD. 
 
Our final results are consistent with the 
target picture for TMD factorisation illustrated in Fig.~\ref{fig:diagrams}: the dijets are produced via photon-gluon fusion and  the gluon TMD  includes one-loop corrections expressing the evolution of the target.
To our knowledge, this is the first time that the DGLAP and CSS evolutions of the gluon distribution in the {\it target} emerges within the CGC approach.  This should be contrasted with the corresponding evolutions for the {\it dilute projectile} (e.g. the proton in $pA$ collisions) or for the {\it fragmentation functions} (partons into hadrons) in the final state, which were already demonstrated in this context~\cite{Dumitru:2005gt,Bergabo:2022tcu,Iancu:2022gpw,Caucal:2024nsb,Altinoluk:2024vgg,Altinoluk:2011qy,Altinoluk:2023hfz}.

The plan of this paper is as follows: Sect.~\ref{sec:1} presents the calculation of the real NLO corrections to back-to-back dijet production in the leading-twist approximation of interest. After a brief review of the respective LO calculation (which also allows us to introduce some notations), we describe the NLO calculation, first in the simpler context  of the Sudakov (virtual and real) logarithms (in Sect.~\ref{sec:Sud}), then in the general case (in Sect.~\ref{sec:NLO}). The NLO results are summarised in Sect.~\ref{sec:NLOTMD}, where we shall also discuss the change of representation, from the CGC picture of projectile evolution to the target picture where the TMD factorisation (here, at NLO) becomes manifest. \eqn{sigmaNLOR}
presents our result for the real NLO corrections to the dijet cross-section. The respective virtual corrections are discussed in Sect.~\ref{sec:beta}, with special emphasis of the single logarithmic piece proportional to $\beta_0$ (the coefficient in the one-loop $\beta$-function), whose emergence in this CGC context is quite subtle. In Sect.~\ref{sec:evol}, the NLO corrections computed previously are promoted into evolution equations: first, the DGLAP equation in Sect.~\ref{sec:DGLAP}, then the (diagonal version of the) CSS equation in transverse momentum space, in Sect.~\ref{sec:CSS}. In particular, we propose a recipe for the matching between the BK/JIMWLK and the DGLAP equations in such a way as to avoid the double-counting of radiative corrections. In the following two sections, we make contact with the standard CSS equations, which truly are a system of two differential equations usually written in transverse coordinate space. In Sect.~\ref{sub:TMD-cs} we perform the Fourier transform of our version of the CSS equation from the transverse momentum  space ($\bK$) to the transverse coordinate space ($\bbp$). Then, in Sect.~\ref{sec:diag} we show that the system of two differential equations encoding the CSS evolution in the general case reduces to our ``diagonal'' equation in the special case where the relevant resolution scales are identified with each other --- as natural in our dijet problem, on physical grounds. In Sect.~\ref{sec:CSSsol} we present an exact but special solution to the ``diagonal'' CSS equation in momentum space, that we compare to numerical solutions obtained via an excursion through the coordinate space. Finally, Sect.~\ref{sec:concl} contains a brief summary and some perspectives.

\section{TMD factorisation for hard dijets beyond leading order}
\label{sec:1}

In this section,  we will  compute the three-parton (quark-antiquark-gluon) component of the light-cone wavefunction (LCWF) of the virtual photon together with the respective contribution to the dijet cross-section in the regime where the transverse momentum imbalance $K_\perp$  of the hard $q\bar q$ pair is controlled by the recoil of the gluon:  $P_\perp \gg K_\perp \simeq  k_{g\perp}\gg Q_s(x)$.  This particular component has been computed in several recent publications for the case of the {\it exact} kinematics
\cite{Taels:2022tza,Caucal:2021ent,Bergabo:2022tcu,Iancu:2022gpw} (see also \cite{Beuf:2011xd,Beuf:2017bpd,Ayala:2017rmh,Ayala:2016lhd} for earlier work and
\cite{Beuf:2022kyp,Iancu:2022lcw,Caucal:2023nci,Caucal:2023fsf,Hauksson:2024bvv}  for related approaches),
yet the simplifications occurring in the regime  of interest for us here have not been systematically studied.
These simplifications are conceptually interesting since they reveal the CSS evolution of the gluon TMD together with the DGLAP evolution of the gluon PDF.

In practice, it turns out that it is more convenient to perform the relevant kinematical approximations already in the early steps of the calculation, rather than trying to simplify the (exact) final results presented in \cite{Taels:2022tza,Iancu:2022gpw}.
Indeed, these approximations are naturally implemented in transverse momentum space and lead
to drastic simplifications of the momentum-space LCWF prior to the final Fourier transform to the transverse
coordinate space (as needed in order to account for multiple scattering in the eikonal approximation). A similar
strategy has been used in \cite{Iancu:2022lcw} in the context of diffractive 2+1 jet production, from where we shall adapt some of the intermediate results.

\subsection{Brief review of the leading order cross-section}
\label{sec:LO}

As a warm-up and also in order to introduce some notations, it is useful to start with a brief review of the
leading-order (LO) result for the dijet cross-section at large $P_\perp\gg K_\perp\gtrsim Q_s(x)$ (``the correlation limit''), which is well known \cite{Dominguez:2010xd}. Here, we shall mostly follow the presentation in \cite{Iancu:2022gpw}, to which we refer for more details.
For convenience, we will only discuss the case of a virtual photon with transverse polarisation ($\gamma_T^*$). The only change which appears for a longitudinal  photon ($\gamma_L^*$) refers to the hard factor, which can be found in the literature \cite{Dominguez:2010xd}.

We construct the LCWF of the virtual photon in the dipole frame, in which the $z$ axis represents the collision axis for the $\gamma^* A$ scattering. The photon is a ultrarelativistic right mover with 4-momentum $q^\mu=(q^+, q^- =- Q^2/2q^+, \bm{0}_\perp)$ and $q^+\gg Q$, while the nuclear target is a ultrarelativistic left mover with 4-momentum $P_N^\mu\simeq \delta^{\mu-}P_N^-$ per nucleon. (We assume $P_N^-\gg M_N$ and hence neglect the nucleon mass $M_N$.)

The quark-antiquark ($q\bar q$) Fock space component of the LCWF has the general structure 
\begin{align}\label{qqLCWF} 
\left|\gamma_{T}^{i}(q)\right\rangle_{q\bar{q}}\, 
= 
\int_{0}^{1}\rmd z_1 \rmd z_2 
\int \frac{\rmd^{2}\bm{k}_1}{(2\pi)^2}  \frac{\rmd^{2}\bm{k}_2}{(2\pi)^2}   \,
\Psi^{i,\,\alpha\beta}_{\lambda_{1}\lambda_{2}}( z_1,  \bk_1,  z_2, \bk_2)
\,
\left|{q}_{\lambda_{1}}^{\alpha}( z_1, \bk_1)\bar{q}_{\lambda_{2}}^{\beta}( z_2, \bk_2) \right\rangle,
\end{align}
while the subscript $i=1,2$ refers to the photon polarisation state, $\lambda_{1,2}$ are helicity states for
the quark and the antiquark, and $\alpha,\,\beta=1\dots N_c$ are colour indices for the fundamental representation. 

In the absence of scattering (i.e. for a vanishing target field $A^-=0$), 
the wavefunction $\Psi^i$ merely describes the photon decay and reads  \cite{Iancu:2022gpw}
\begin{align}
	\label{psiqqnew}
	\Psi^{i,\,\alpha\beta}_{\lambda_1 \lambda_{2}} ( z_1,  \bk_1,  z_2, \bk_2)\Big |_{A^-=0}
=  (2\pi)^2\delta^{(2)}(\bm{k}_1 +\bm{k}_2)\,\delta(1\minus z_1 \minus  z_2)
	\delta_{\lambda_1\lambda_2}\,\delta^{\alpha\beta}\,
	\sqrt{\frac{q^+}{2}}\,
	\frac{ee_{f}}{2\pi}\,
	\frac{\varphi^{ij}(z_1,\lambda_1)\,k_1^{j}}
	{k_{1\perp}^2 +\bar{Q}^2},
\end{align}
where $e$ is the QED coupling, $e_f$ the fractional electric charge of the quark with flavor $f$,
 $\bar{Q}^2 \equiv z_1 z_2 Q^2$ and  the function
\begin{align}
	\label{phidef}
	\varphi^{ij}(z,\lambda)
	\equiv
	(2z-1)\delta^{ij}
	+ 2 i \lambda \varepsilon^{ij}
\end{align}
with $\varepsilon^{ij}$ the Levi-Civita symbol in two dimensions, encodes the helicity structure of the photon splitting vertex. 

The high energy scattering (as computed in the eikonal approximation) does not modify the
longitudinal momenta of the two fermions, nor their helicities, but it can change their transverse momenta
and it rotates their colour states. These effects are most conveniently implemented in
the transverse coordinate representation. Observing that
\beq\label{eq:int}
(2\pi)^2\delta^{(2)}(\bm{k}_1 +\bm{k}_2)\,\frac{k_1^j}{k_{1\perp}^2 +\bar{Q}^2}
= \frac{i}{2\pi}
	\int \dif^2 \bx  \dif^2 \by \, e^{ - i \bk_1 \cdot \bx - i \bk_2 \cdot \by}\,
	\frac{x^j-y^j}{|\bx-\by|}\,
	\bar{Q} K_1(\bar{Q} |\bx-\by|),
	\eeq
it should be clear that the effect of the collision amounts to inserting the colour matrix 
$(V_{\bx} V_{\by}^\dagger)^{\alpha\beta}$ inside the integrand in \eqn{eq:int}. Here $V_{\bx}$ 
and $V_{\by}^\dagger$ are Wilson lines describing the colour precessions of the quark and the antiquark, e.g.
 \begin{align}
	\label{wilson}
	V(\bm{x})={\rm T}
	\exp\left[i g \int \rmd x^{+} 
	t^{a} A^{-}_a(x^{+},\bm{x})\right],
\end{align}
where the  symbol T stands for LC time ($x^+$) ordering and $t^a$ are the generators of SU$(N_c)$ in the
fundamental representation. We omit the $\perp$ symbol on the transverse coordinates to keep notations simple (e.g.~$\bm{x} \equiv \bm{x}_\perp$).

At this point, we take into account the fact that the two fermions are hard and nearly back to back, meaning that they scatter only weakly. To that aim, it is convenient to introduce relative and central coordinates together with their conjugate momenta: 
 \begin{align}
	\label{pandk}
	\br\equiv \bx-\by\,,
	\qquad \bb\equiv z_1\bx+z_2\by\,,	\qquad 
	\bP \equiv z_2\bk_1 - 
	z_1 \bk_2\,,
	\qquad
	\bK \equiv \bk_1 + \bk_2\,,
\end{align}
so that  $\bk_1 \cdot \bx + \bk_2 \cdot \by=\bP \cdot \br +\bK \cdot \bb$.  The hard regime (or ``correlation limit'')
corresponds to $P_\perp\gg K_\perp \gtrsim Q_s(x)$, which in turn implies that the dipole transverse size $r\sim 1/P_\perp$ is much smaller than the typical distance $1/Q_s(x)$ for spatial variations in the target gluon distribution. We can therefore expand the $S$--matrix as
\begin{align}
	\label{Vqq}
	V_{\bx} V_{\by}^\dagger - 1\simeq -r^j V_{\bb}\partial^j V_{\bb}^\dagger.
	\end{align}
This holds up to corrections suppressed by powers of $K_\perp/P_\perp$ (``kinematic twists'')
and/or by powers of $Q_s/P_\perp$ (saturation effects). The factor $r^j$ in
the final result is the hallmark of the dipolar coupling, while the colour operator $\mcal{A}^i(\bb)\equiv
(i/g)V_{\bb}\partial^j V_{\bb}^\dagger\equiv \mcal{A}^i_a(\bb) t^a$ 
is recognised as the colour field of the target in the {\it target} light-cone gauge $\mcal{A}^-=0$ \cite{Iancu:2002xk,Iancu:2003xm}. Importantly, \eqn{Vqq} shows that the dependences of the scattering matrix
upon the variables $\br$ and $\bb$ {\it factorise} from each other, which together with \eqn{eq:int} implies
a similar factorisation in the transverse momentum variables $\bP$ and $\bK$: this is the premise of TMD factorisation for the process at hand. Specifically, after changing integration variables ($\bx,,\,\bx\,\to\,\br,\,\bb$) in  \eqn{eq:int} and performing the integral over $\br$, one is left with
\begin{align}\label{PsiLO}
	\Psi^i_{\lambda_1\lambda_2}(z_1, \bP, \bK)
	&\, = -i \sqrt{\frac{q^+}{2}}\,\frac{ee_{f}}{2\pi}\,\delta_{\lambda_1\lambda_2}
 {\varphi^{il}(z_1,\lambda_1)}
	\mcal{H}^{lj}(z_1, \bP)
	\int \rmd^2\bb\,e^{-i \bm{K}_\perp\cdot \bb} 
	\big(V_{\bb}\partial^j V_{\bb}^\dagger\big),
	\end{align}
where the colour indices and the $\delta$-function $\delta(1\minus z_1 \minus  z_2)$ are kept implicit and the ``hard'' tensorial factor
\begin{align}
	\label{hardamp}
	\mcal{H}^{lj}(z_1, \bP) \equiv
	\int 
	\frac{\rmd^2 \br}{2\pi}\,
	e^{-i \bP \cdot \br}
	\frac{r^l r^j}{r}\,
	\bar{Q} K_1(\bar{Q}r) =
	\frac{1}{P_\perp^2 + \bar{Q}^2}
	\left(
	\delta^{lj} - 
	\frac{2 P^l P^j}{P_{\perp}^2+\bar{Q}^2}
	\right)
\end{align}
describes the $P_\perp$--distribution of the $q\bar q$ pair, as generated by the 
photon decay followed by the dipolar coupling to the field of the target (cf. \eqn{Vqq}).
The final factor in \eqn{PsiLO} --- the Fourier transform of the target field --- makes  it clear
that the dijet imbalance $\bK$ is controlled by the scattering.

\eqn{PsiLO} is our final result for the $q\bar q$ wavefunction in the leading twist approximation at large
$P_\perp$. Via standard manipulations, it is now easy to deduce the following expression for the dijet
cross-section~\cite{Dominguez:2010xd,Dominguez:2011wm}
\begin{align}
	\label{sigma0}
	\frac{\dif \sigma^{\gamma_{T}^* A \to q\bar{q}X}_\lo}
	{\rmd z_1
  	\rmd z_2\dif^2\bP\dif^2\bK } &\,=  {\alpha_{em}\alpha_s}
	e_f^2\delta(1-z_1-z_2)\left(z_1^{2} + 
	z_2^{2}\right) \frac{P_{\perp}^4 + \bar{Q}^4}
	{(P_{\perp}^2 + \bar{Q}^2)^4}\nonumber\\*[0.2cm]
	&\, \times \left\{\,\mcal{F}_g^{(0)}(x, K_{\perp})
	 -   \frac{4 P_{\perp}^2 \bar{Q}^2}{P_{\perp}^4 + \bar{Q}^4}
        \left[ \frac{( \bP\cdot \bK)^2}{P_{\perp}^2\,K_{\perp}^2 }
       -\frac{1}{2}
        \right]  \mcal{H}_g^{(0)}(x,K_{\perp})\right\},
\end{align}
where $\mcal{F}_g^{(0)}(x, K_{\perp})$ and $\mcal{H}_g^{(0)}(x,K_{\perp})$ are the (leading-order) unpolarised and respectively 
linearly polarised pieces of the tensorial Weiszäcker-Williams (WW) gluon TMD, as defined via the following
 decomposition of the Fourier transform of the 2-point function of the transverse fields $\mcal{A}^i(\bb)$  \cite{Metz:2011wb}
 \begin{align}
 \hspace*{-0.2cm}
 	\label{dGWW} \mcal{W}^{ij}(x, \bK)
	&\ \equiv  \,\frac{1}{(2\pi)^4}
  \int\rmd^2{\bm{b}}\, \rmd^2 \bbb \ 
   \rme^{-i\bK\cdot(\bb-\bbb)}\   \frac{-2}{\alpha_s}
\left\langle \! {\rm Tr} \!\left[ 
V_{\bb} (\del^i V_{\bb}^{\dagger})\,
     V_{\bbb} (\del^j V_{\bbb}^{\dagger})\right]\right\rangle_{x}
     \nonumber\\*[0.2cm]
	&\ \equiv \,\frac{\delta^{ij}}{2}\,
\mcal{F}_g^{(0)}(x, \bK) +
        \left(\frac{K_{\perp}^i K_{\perp}^j}{K_{\perp}^2}
        -\frac{\delta^{ij}}{2} \right)
        \mcal{H}_g^{(0)}(x,\bK).
   \end{align}
 The target expectation value is defined in the sense of the CGC effective theory, that is,
 as an average over the colour fields $A^-_a$ in the projectile LC gauge. It is understood that
 the CGC weight function is evaluated at the value $x=x_{q\bar q}$, where $x_{q\bar q} P_N^-$ is
 the target longitudinal momentum which is transferred to the $q\bar q$ pair via the collision. In turn, this is
determined by the condition that the final $q\bar q$ pair be on-shell, that is (with $\hat s\equiv 2q^+P_N^-$):
\beq\label{xqqdef}
 x_{q\bar q} = \frac{1}{\hat s}\left(Q^2+\frac{k^2_{1\perp}}{z_1} +\frac{k^2_{2\perp}}{z_2} \right)
  \simeq\, \frac{1}{\hat s}\left(Q^2+\frac{P_\perp^2}{z_1 z_2}\right).
  \eeq
As manifest in \eqn{sigma0}, the linearly polarised piece of the TMD measures the strength
   of the angular correlation between the transverse vectors $\bP$ and $\bK$. If one integrates out $\bK$, i.e. if one is not interested in measuring the dijet imbalance, this angular correlation averages out to zero and we are left with the standard result for the dijet cross-section in collinear factorisation, which involves the gluon PDF at the scale $P_\perp^2\sim Q^2$:
   \begin{align}
	\label{sigma0int}
	\frac{\dif \sigma^{\gamma_{T}^* A \to q\bar{q}X}_\lo}
	{\rmd z_1
  	\rmd z_2\dif^2\bP}=  {\alpha_{em}\alpha_s}
	e_f^2\delta(1-z_1-z_2)\left(z_1^{2} + 
	z_2^{2}\right) 
	\frac{P_{\perp}^4 + \bar{Q}^4}
	{(P_{\perp}^2 + \bar{Q}^2)^4}\,xG^{(0)}(x, P^2_{\perp}).
\end{align}
with the LO gluon PDF:
 \beq\label{PDF0}
xG^{(0)} (x, \PT^2) =\pi \int^{P_\perp^2}_{\Lambda^2}\!{\rmd \ell_\perp^2}
\mcal{F}_g^{(0)}\left( x, \ell_{\perp}\right),\eeq
where $\Lambda$ is an infrared cutoff of the order of the confinement scale.

As well documented in the literature (see e.g. \cite{McLerran:1998nk,Mueller:1999wm,Iancu:2002xk,Iancu:2003xm,Dominguez:2010xd}), the unpolarised WW TMD coincides with the unintegrated gluon distribution at small $x$. Its properties have been studied {\it in extenso} within the CGC effective theory (see e.g.  \cite{Iancu:2002xk,Iancu:2003xm} for pedagogical discussions using the McLerran-Venugopalan model~\cite{McLerran:1993ni,McLerran:1993ka} and  \cite{Marquet:2016cgx,Marquet:2017xwy} for exhaustive numerical studies which include the JIMWLK evolution). For what follows, it suffices to recall that there are two main physical regimes, depending upon the ratio $K_\perp/Q_s(x)$: \texttt{(i)} in the saturation regime at relatively low transverse momenta $K_\perp\lesssim Q_s(x)$, the unintegrated gluon distribution is controlled by multiple soft scattering and it exhibits a saturation plateau with a height of order $1/\alpha_s$, which increases logarithmically with decreasing $K_\perp$ below $Q_s(x)$;  \texttt{(ii)} for much larger momenta $K_\perp\gg Q_s(x)$, the unintegrated gluon distribution is determined by a single hard scattering, which generates a typical pQCD power tail in $1/K_\perp^2$. The change in this tail at large $K_\perp$ is the main scope of the quantum (DGLAP+CSS) evolution to be discussed in the next sections.

 \subsection{Towards NLO: the Sudakov logarithms as a warm-up}
   \label{sec:Sud}

When going beyond (strict) LO, one expects the gluon PDF $xG^{(0)} (x,\PT^2)$ which appears in  the  cross-section in \eqn{sigma0int} to obey the DGLAP evolution with increasing $\PT^2$. Whereas the emergence of this evolution is well understood in the context of the collinear factorisation in the target picture,  it is {\it a priori} less clear whether this is also encoded in the CGC framework. In what follows, we would like to demonstrate that this is indeed case  via an explicit study of the NLO corrections to dijet production in the dipole picture.

 To that aim, we consider (real and virtual) gluon emissions by the $q\bar q$ pair. As we shall see, the emissions
 responsible for the DGLAP dynamics in this  picture have  transverse momenta $k_{g\perp}$
  in the range $K_\perp \le k_{g\perp} \ll P_\perp$ and longitudinal momenta $k_g^+=z_g q^+$
 with $\kg^2/P_\perp^2 \lesssim z_g \lesssim
\kg/P_\perp\!\ll\! 1$. The lower limit on $z_g$ is the boundary between the  {\it very} soft emissions with lifetime $\tau_g\equiv 2z_qq^+/k_{g\perp}^2$ much shorter than the
coherence time $\tau_\gamma=2q^+/Q^2$ of the virtual photon, which must be
included in the BK/JIMWLK evolution \cite{Balitsky:1995ub,Kovchegov:1999yj,JalilianMarian:1997jx,JalilianMarian:1997gr,Kovner:2000pt,Weigert:2000gi,Iancu:2000hn,Iancu:2001ad,Ferreiro:2001qy} with collinear improvement
 \cite{Beuf:2014uia,Iancu:2015vea,Hatta:2016ujq,Ducloue:2019ezk} (see also the discussion in Sect.~\ref{sec:NLO} below),
and the {\it less} soft emissions ($\tau_g\gtrsim \tau_\gamma$), as relevant for the DGLAP dynamics.

To understand the other kinematical constraints on the gluon emissions, it is instructive to first consider the case of a {\it final-state} emission, which occurs after the collision ($\tau_g\gg \tau_\gamma$), see Fig.~\ref{fig:gluon_after}. Final-state emissions of soft gluons are well known to factorise, thereby preserving the TMD factorisation found at LO, cf. \eqn{sigma0}. 
To start with, consider a gluon which is both emitted and reabsorbed by the quark. To the approximations of interest, its  effect on the cross-section can be seen as an additional contribution  to the gluon TMD (we consider only the unpolarised piece, for simplicity):
  \begin{align}
\label{Frealfin}\hspace*{-0.2cm}
 \Delta \mcal{F}^{qq}_{\rm fin}(x_{q\bar q}, \bK)&=
    \int \rmd^2\bk_g \int_{\kg^2/\PT^2}^{1}\frac{\rmd z_g}{z_g} 
 \frac{1}{\left(\bk_g-\frac{z_g}{z_1}\bP\right)^2}\nonumber\\
 &\hspace{-0.5cm}\times\frac{\alpha_s C_F}{\pi^2}\left[ \mcal{F}_g^{(0)}(x_{q\bar q g},\bK+\bk_{g\perp})-\mcal{F}_g^{(0)} (x_{q\bar q},\bK)\right].
\end{align}
The two terms within the square brackets refer to real emissions and self-energy corrections, respectively. 
They are evaluated at different values of $x$ (the longitudinal momentum fraction w.r.t. $P_N^-$), as they correspond to different final state.  For the virtual term, we have a quark-antiquark final state, so the relevant value of $x$ is $x_{q\bar q}$, as defined in \eqn{xqqdef}. For the real term, the final state also includes a gluon; the condition that the three parton system ($q\bar q g$) be on-shell determines the ``minus'' longitudinal momentum fraction that must be transferred by the target:
\beq\label{xgdef}
  x_{q\bar q g} =\frac{1}{2q^+P_N^-}\left(Q^2+\frac{k^2_{1\perp}}{z_1} +\frac{k^2_{2\perp}}{z_2} 
 +\frac{k^2_{g\perp}}{z_g}\right)\simeq\,\frac{1}{\hat s}\left(Q^2+\frac{P^2_{\perp}}{z_1z_2} 
  +\frac{k^2_{g\perp}}{z_g}\right).\eeq
 We have anticipated here the fact that $z_g\ll 1$, hence $z_1+z_2\simeq 1$, as at LO.

The denominator  in \eqn{Frealfin} exhibits a collinear singularity when  the relative transverse momentum of the quark-gluon pair $z_1\bk_{g\perp}- {z_g}\bP$ approaches to zero.  Here however we are interested in gluon emissions at large angles, outside the jet generated by the quark, which are not concerned by this singularity.
Such large-angle emissions are interesting for at least two reasons: \texttt{(i)} they modify the cross-section for  $q\bar q$ dijet production, and \texttt{(ii)} they have a phase-space which is  logarithmic in both transverse ($k_{g\perp}$) and longitudinal ($z_g$) momenta, so they yield a large NLO contribution to the dijet cross-section, which is enhanced by a double logarithmic (the ``Sudadov double log'').

To see this, notice that the gluon propagation can be estimated as $\theta_g\simeq \kg/(z_gq^+)$. The gluon is emitted outside the quark jet provided this angle is much larger than the  propagation angle $\theta_q\simeq  P_\perp/q^+$ of the quark. The condition $\theta_g \gg  \kg/P_\perp$ implies $z_g \ll \kg/P_\perp$, meaning that the denominator  in \eqn{Frealfin} can be approximated as
\beq
\left(\bk_g-\frac{z_g}{z_1}\bP\right)^2\,\simeq\, \kg^2\,,\eeq
which is independent of $z_g$. Accordingly, both integrations in \eqn{Frealfin} become logarithmic, as anticipated. The above argument also shows that the integral over $z_g$ must be limited to $z_g < \kg/P_\perp$.  To the double-logarithmic accuracy of interest, one can ignore the difference between the longitudinal fractions in the real and the virtual term, respectively: indeed, we are interested in relatively large values $z_g \gg \kg^2/\PT^2$, for which one can approximate $x_{q\bar q g}\simeq x_{q\bar q}\equiv x$.

Consider now the boundaries of the transverse intergation.
 For low transverse momenta $\kg\ll\KT$,  the real and virtual contributions to Eq.\,\eqref{Frealfin} mutually cancel, since $\mcal{F}_g(x,\bK+\bk_g)\simeq \mcal{F}_g(x,\bK)$. For larger momenta $\kg\gg\KT$, the real corrections are power suppressed since $\mcal{F}_g(x,\bK+\bk_g)\simeq \mcal{F}_g(x,\kg)\propto 1/\kg^2$. Concerning the virtual emissions, the integral over $\kg$ is logarithmic and exhibits an ultraviolet divergence, that  cancels out after adding all the NLO contributions and gets
replaced by an upper cutoff of order $\PT$
\cite{Caucal:2021ent,Taels:2022tza,Caucal:2022ulg,Caucal:2023nci,Caucal:2023fsf}.

Thus, the net result for  $\Delta \mcal{F}^{qq}_{\rm Sud}$  to double-logarithmic accuracy (DLA) 
comes from virtual emissions with transverse momenta $\KT\ll \kg\ll\PT$ and longitudinal 
fractions $\kg^2/P_\perp^2 \ll z_g \ll \kg/P_\perp$, and reads 
  \begin{align}
\label{Fqq}\hspace*{-0.1cm}
 \Delta \mcal{F}^{qq}_{\rm Sud}=
 -\frac{\alpha_s C_F}{\pi}\, \mcal{F}_g^{(0)} (x,\bK)
    \int_{\KT^2}^{\PT^2}\frac{\rmd\kg^2}{\kg^2} \int_{\frac{\kg^2}{\PT^2}}^{\frac{\kg}{\PT}}\frac{\rmd z_g}{z_g}.
\end{align}
Clearly, there is an identical contributions from the direct emissions  by the antiquark. For such large angle emissions, one also needs to consider the interference terms (gluon exchanges between $q$ and $\bar q$), whose leading twist contribution at large $P_\perp$  is of order $1/N_c$. The overall effect of the final-state gluon emissions to DLA is given by an expression similar
to \eqn{Fqq}, but with the colour factor $C_F$ replaced by $N_c=2C_F+ 1/N_c$:
\begin{align}
\label{FSukvirt}
 \Delta \mcal{F}^{\mcal{V}}_{\rm Sud}(x, \bK,\PT^2)   
  =-\frac{\alpha_s N_c}{4\pi}\,\ln^2\frac{\PT^2}{\KT^2}\,\mcal{F}_g^{(0)} (x,\bK)\,.
\end{align}
This is recognised as the Sudakov double logarithm for dijet production  in DIS
\cite{Mueller:2013wwa,Taels:2022tza,Caucal:2022ulg}. As shown by the notation, this also depends upon the hard scale $P_\perp$, which acts as a resolution scale for measuring the gluon emissions. Importantly, this hard {\it transverse} momentum controls {\it both} the transverse and the longitudinal resolutions: it plays the role of an UV cutoff on  $\kg$ and also enters the lower and upper limits of the integral over $z_g$.

\comment{The above discussion also shows that the phase-space for soft ($z_g\ll 1$) gluon emissions is not fully covered by the BK/JIMWLK evolution: when $\kg\ll \PT$, this evolution is restricted to smaller values $z_g\ll \kg^2/\PT^2$. The remaining phase-space at $z_g\gtrsim \kg^2/\PT^2 $ corresponds to late time emissions, that can be either outside the jets (so long as $z_g \lesssim \kg/\PT$), or inside them (for 
 $\kg/\PT<  z_g \ll 1$). The intra-jet emissions do not matter for the dijet cross-section (but only for the jet substructure), so they will not be considered in what follows. Rather we shall focus on the emissions in the intermediate range at $\kg^2/\PT^2 \lesssim z_g \lesssim \kg/\PT$ which, as we shall demonstrate, generate the DGLAP and CSS evolutions (with the latter encompassing the Sudakov effects).
 }
 
 The upper index $\mcal{V}$ on the  Sudakov double log in \eqn{FSukvirt} is meant to emphasise that the {\it net} contribution comes from {\it uncompensated virtual emissions}. But of course, the real emissions too were important in order to obtain this result: as shown by the previous discussion, they have limited the phase-space for the logarithmic integrations.

Here comes our first new observation: if the imbalance $\KT$ itself is relatively hard, 
$Q_s \ll \KT\ll \PT$, there is an additional Sudakov effect, coming from the {\it real}
emissions with $\kg\simeq \KT$. The dijet imbalance is then controlled by the gluon recoil: $\bK\simeq -\bk_g$.  A similar effect has been noticed  for  $pA$ collisions in Ref.~\cite{Altinoluk:2011qy,Altinoluk:2023hfz}, but in that case the gluon emission is hard and contributes to the $s$-channel  DGLAP/CSS evolution of the {\it projectile} (the proton), or of the fragmentation functions (for hadron production)~\cite{Caucal:2024nsb,Dumitru:2005gt,Altinoluk:2024vgg}. In our case, the gluon is still soft compared to its parent ($z_g \ll 1$ and $ \KT\ll \PT$) and contributes to the $t$-channel evolution of the {\it target} (the nucleus), as we shall later argue.

Specifically, consider the real term in Eq.\,\eqref{Frealfin} in
the regime where $\ell_\perp\ll \KT$, with $\bm{\ell}\equiv \bK+\bk_g$
the transverse momentum transferred by the target via the collision. After changing the integration variable
from $\bk_g$ to $\ellt$ and adding the  interference terms, one finds
\begin{align}
\label{FSukreal}
 \Delta \mcal{F}^{\mcal{R}}_{\rm Sud} (x, \bK, \PT^2)   =
 \frac{\alpha_s N_c}{\pi^2}\,\frac{1}{\KT^2} \int_{\frac{\KT^2}{\PT^2}}^{\frac{\KT}{\PT}}\frac{\rmd z_g}{z_g} 
 \int^{\KT^2}_{\Lambda^2} \rmd^2\ellt\, \mcal{F}^{(0)}_g(x,\ellt)
  = \frac{\alpha_s N_c}{2\pi^2}\,\frac{1}{\KT^2} \ln\frac{\PT^2}{\KT^2}\,xG^{(0)} (x,\KT^2).
\end{align} 
The longitudinal integration has generated the Sudakov logarithm, while that 
over $\ell_\perp$ has built the gluon PDF, as shown in the second equality.
The integral over $\ell_\perp$  is logarithmic too when $\ell_\perp\gg Q_s$,
so $xG^{(0)} (x,\KT^2)\propto \ln(\KT^2/Q_s^2)$.  So, the ``real'' Sudakov in \eqn{FSukreal} is proportional
too to a double logarithm, albeit this is not the same as that from the virtual term.
The virtual Sudakov \eqn{FSukvirt} dominates when
$\KT$ is relatively low, of order $Q_s$, yet both contributions --- real and virtual ---  behave like $1/\KT^2$ when $\KT\gg Q_s$, so they
contribute on the same footing when integrating over $\KT$ up to $\PT$. After this integration, the real and virtual Sudakov effects mutually cancel:
\beq
\int^{\KT^2}_{\Lambda^2}\rmd^2\bK
 \left\{  \Delta \mcal{F}^{\mcal{V}}_{\rm Sud}(x, \bK,\PT^2)   + \Delta \mcal{F}^{\mcal{R}}_{\rm Sud} (x, \bK, \PT^2) 
 \right\}=0,\eeq
as one can easily check. Such a cancellation was  expected on physical grounds \cite{Mueller:2013wwa}, yet it has not been verified in previous studies which only considered the virtual Sudakov~\cite{Mueller:2013wwa,Taels:2022tza,Caucal:2022ulg}. This cancellation also shows that the total number of gluons $xG^{(0)}(x,\PT^2)$ is not modified by the Sudakov dynamics. To  uncover the DGLAP evolution,  one must  push the NLO calculation beyond DLA.

To that aim, one needs a complete calculation of the NLO corrections in the kinematical regime of interest,
namely, $P_\perp\gg K_\perp\gg \lt\sim Q_s(x)$ and  $K_\perp^2/P_\perp^2 \lesssim z_g \lesssim K_\perp/P_\perp$. That is, beside the Sudakov effects already computed, one also needs the respective contributions from gluon emissions which occur {\it before} the scattering (``initial-state''). These contributions must be computed in a ``leading-twist'' approximations in which one systematically keeps the leading order terms in the double expansion in powers of $K_\perp/\PT$ and $\ell_\perp/K_\perp$.

\subsection{The NLO dijet amplitude at leading twist}
\label{sec:NLO}

In this section, we will demonstrate the consistency between TMD factorisation and the NLO corrections to dijet production, as computed in the CGC effective theory at small $x$. More precisely, we would like to demonstrate that these corrections preserve TMD factorisation and generate the expected DGLAP+CSS evolution of the gluon TMD --- in agreement with the traditional  one-loop calculations in the target picture and at moderate $x$~\cite{Collins:2011zzd,Boussarie:2023izj}. To that aim one needs to consider one gluon emission (real or virtual) with a special kinematics: 

 \texttt{(1)} the gluon transverse momentum $\kg$ must be {\it relatively large} --- much smaller than the dijet relative momentum $P_\perp$, but much larger than the momentum $\lt\sim Q_s(x)$ transferred by the target. This ensures that the gluon emission controls the dijet momentum imbalance, 
$K_\perp \simeq  k_{g\perp}$,  and thus it renormalises the tail of the gluon TMD at high $K_\perp$.
The change in that tail is the hallmark of the DGLAP dynamics.

\texttt{(2)} the gluon must be {\it relatively soft} w.r.t. to the photon, that is, its ``plus'' longitudinal momentum fraction $z_g$ must be small, but not {\it too} small, to avoid the overlap with the {\it very} soft gluon emissions which are responsible for the high-energy evolution of the WW gluon TMD  \cite{Dominguez:2011gc}. As anticipated in the previous section, the relevant condition reads
\beq\label{zg}
x_*\,\frac{\kg^2}{P_\perp^2}\, \le\, z_g \,\lesssim\, \frac{\kg}{P_\perp}\,\ll\,1\,,\eeq
where the lower limit is introduced by the separation from the high-energy evolution with collinear improvement \cite{Beuf:2014uia,Iancu:2015vea,Hatta:2016ujq,Ducloue:2019ezk} (see also the discussion in \cite{Taels:2022tza,Caucal:2023nci}), while the upper limit is the condition that the gluon  be emitted at large angle, outside the dijets. (For di-hadron production, this upper limit would be replaced by 1.)

As compared to the discussion in Sect.~\ref{sec:Sud}, the lower limit in \eqn{zg} includes an additional factor $x_\star$, which is needed in order to more precisely separate the longitudinal phase-space between the high-energy and the collinear evolutions.
This is a pure number obeying  $x_\star\ll 1$ and  $\alpha_s\ln(1/x_\star)\ll1$.
That is, the rapidity scale for evaluating the high-energy evolution of the WW gluon TMD in the LO cross-section  \eqn{sigma0} should be $x_{q\bar q}/x_*$, and not $x_{q\bar q}$. The difference between these two scales may seem irrelevant from the viewpoint of the high energy evolution, since $\alpha_s\ln(1/x_\star)\ll 1$. Yet, in the presence of widely separated transverse scales, this difference can be enhanced by the transverse logarithm $\ln(\KT^2/Q_s^2)$, thus yielding an effect of $\order{1}$. We shall later demonstrate that such $\order{1}$ effects cancel between the BK/JIMWLK evolution of the gluon TMD and the DGLAP evolution of the gluon PDF.

In this section, we consider real emissions, that is, we consider the $q\bar q g$ component of the photon LCWF, which has the following general structure:
\begin{align}\label{qqgLCWF} 
\left|\gamma_{T}^{ij}(q)\right\rangle_{q\bar{q}g}\, 
&\,=  \int_{0}^{1}\rmd z_1 \rmd z_2  \rmd z_g \,\delta(1-z_1-z_2-z_g)\,
\int \frac{\rmd^{2}\bm{k}_1}{(2\pi)^2}  \frac{\rmd^{2}\bm{k}_2}{(2\pi)^2} 
 \frac{\rmd^{2}\bm{k}_g}{(2\pi)^2}  
 \nn*[0.2cm] &\,\times\, \,\delta_{\lambda_1\lambda_2}
\Psi^{ija}_{\alpha\beta}(\lambda_1, z_1, z_g, \bk_1,  \bk_2, \bk_g)
\,
\left|{q}_{\lambda_{1}}^{\alpha}( z_1, \bk_1)\bar{q}_{\lambda_{2}}^{\beta}( z_2, \bk_2) g_j^a( z_3, \bk_g)\right\rangle.
\end{align}
In writing this expression, we have already used the fact that the high energy scattering does not change the ``plus'' longitudinal momenta of the three partons, nor the quark helicities. As compared to the LO result in \eqn{qqLCWF}, we now have two additional external indices $j=1,2$ and $a=1,\dots N_c^2-1$, which refer to the gluon (transverse) polarisation and colour, respectively. Since $z_g\ll z_{1,2}$, one can neglect $z_g$ in the $\delta$-function for longitudinal momentum conservation, but it is important to keep the $z_g$ dependence of the amplitude to the accuracy of interest. Let us also list the various linear combinations of transverse momenta and coordinates
that will be useful in what follows:
\begin{align}\label{PKL}
\bP &\, \equiv \frac{z_2\bk_1- z_1\bk_2}{z_1+z_2}\simeq z_2\bk_1- z_1\bk_2,\qquad
\bK \equiv \bk_1 + \bk_2,\qquad\ellt\equiv \bk_1 + \bk_2+\bkg=\bK+\bkg,
 \nn*[0.2cm] &\,\br\equiv \bx-\by\,,
	\qquad \bb\equiv \frac{z_1\bx+z_2\by}{z_1+z_2}\,\simeq z_1\bx+z_2\by
	\,,	\qquad \bR\equiv \bz-\bb\,,
\end{align}
which implies
\beq
\bk_1 \cdot \bx + \bk_2 \cdot \by + \bk_g \cdot \bz=\bP \cdot \br +\bK \cdot \bb +  \bk_g \cdot \bz
= \bP \cdot \br -\bK \cdot \bR +  \ellt \cdot \bz.\eeq

At this point, it is important to observe that, despite the fact that $z_g\ll 1$, one cannot compute the gluon emission within the eikonal approximation (say, as used for the high-energy evolution). This can be easily understood by recalling the expression \eqn{xgdef} for the ``minus'' longitudinal momentum fraction that must be transferred by the target in order to put the three final partons on-shell.
For the very soft gluons with $z_g \ll k^2_{g\perp}/Q^2$, it is enough the keep the last term inside the parentheses in \eqn{xgdef}, that is, the gluon LC energy $k_g^-={k^2_{g\perp}}/{(2z_gq^+)}$. Similar approximations can be made inside the energy denominators and the gluon emission vertex, thus eventually generating the logarithmic phase-space responsible for the high-energy evolution:
\beq\label{smallx}
\alpha_s\int_{k^2_{g\perp}/\hat s}^{x_*k^2_{g\perp}/Q^2}\frac{\rmd z_g}{z_g}\,=\,
\alpha_s \ln\frac{x_*}{x}\,.\eeq
Here $x=Q^2/\hat s$ and the lower limit on $z_g$ comes from the condition that the gluon formation time $\tau_g\equiv 2z_gq^+/k_{g\perp}^2$ be (much) larger than the longitudinal extent $\sim 1/P_N^-$ of the target.
However, for the larger values of $z_g$ of interest for us here, cf. \eqn{zg}, all the terms within the parentheses in \eqn{xgdef} are generally of the same order, so none of them can be neglected. A similar comment applies to the energy denominators and to the gluon emission vertex. And indeed, the ensuing integrations over $z_g$ are generally {\it not} logarithmic --- with the noticeable exception of the gluon emissions occurring in the final state (after the scattering).

\begin{figure}[t]
    \centering
    \includegraphics[width=0.8\textwidth]{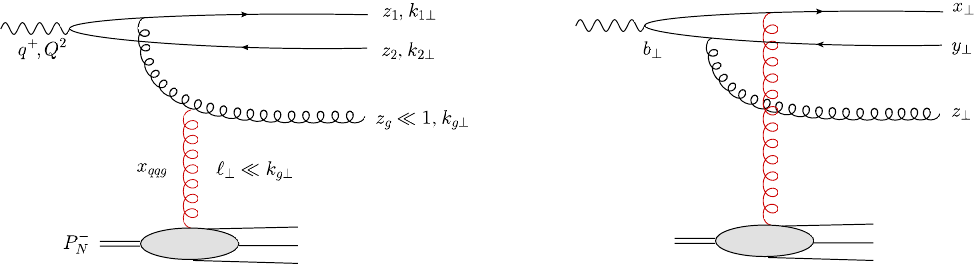}
    \caption{Feynman diagrams representing the quark-antiquark-gluon Fock space component of the virtual photon  in the dipole picture. The gluon is emitted before the scattering with the nuclear target, by either the quark, or the antiquark. The scattering is depicted in the one-gluon exchange approximation, for simplicity. The drawing is meant to be suggestive of the transverse geometry of the collision in the kinematical conditions of interest: the gluon is soft ($z_g\ll 1$) and semi-hard ($Q_s\ll k_{g\perp}\ll k_{1\perp}\sim k_{2\perp}$).}
    \label{fig:3jets_B}
\end{figure}

\begin{figure}[t]
    \centering
    \includegraphics[width=0.8\textwidth]{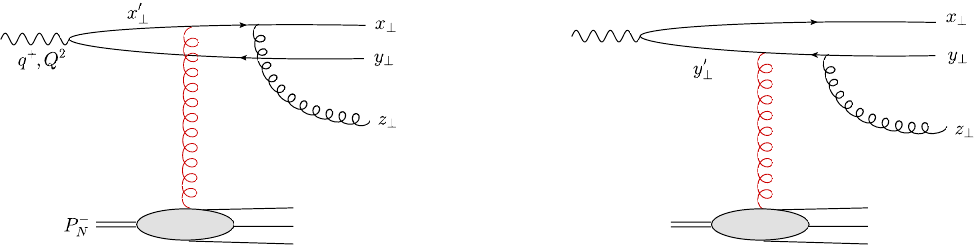}
    \caption{The same as in the previous figure, but for a gluon emission occurring after the  scattering with the nuclear target.}
    \label{fig:3jets_A}
\end{figure}

The diagrams which contribute to the $q\bar q g$ final state in light-cone perturbation theory are shown in
Figs.~\ref{fig:3jets_B} and \ref{fig:3jets_A}. They describe gluon emissions by either the quark, or the antiquark, which occur either before (Fig.~\ref{fig:3jets_B}), or after (Fig.~\ref{fig:3jets_A}), the scattering with the nuclear shockwave. These diagrams must be computed in the projectile light-cone gauge $A^+_a=0$. In this singular gauge, there are two additional diagrams involving the instantaneous 
quark propagator. Although not explicitly shown, these graphs  are important too for the calculation\footnote{They would be negligible in the eikonal approximation at very small $z_g$.}: they must be combined
with the gluon emissions prior to the shockwave to provide gauge-invariant results \cite{Iancu:2022lcw}.

\subsubsection{Gluon emission before the collision}

We start with the four diagrams describing gluon emissions prior to the collision: the two graphs shown in Fig.~\ref{fig:3jets_B} plus the  instantaneous graphs. Those graphs have already been computed in  Ref.~\cite{Iancu:2022lcw} for a similar kinematics (a hard $q\bar q$ pair together with a soft gluon), but in a different physical context: the diffractive production of 2+1 jets. This difference affects the nature of the scattering with the target (elastic in Ref.~\cite{Iancu:2022lcw} versus inelastic in the problem at hand), but not also the structure of the LCWF in the absence of scattering, that we can directly borrow from Ref.~\cite{Iancu:2022lcw}. It reads
\begin{align}\label{Psig0}
	\Psi^{ija}_{\alpha\beta}(\lambda_1, z_1, z_g,\bP, \bK,\bk_{g\perp})\Big |_{A^-=0}
	&\, =	 (2\pi)^2\delta^{(2)}(\bK +\bm{k}_{g\perp})\,t^a_{\alpha\beta}\,
	\frac{e e_f g  q^+}{(2\pi)^2}\,\frac{\varphi^{il}(z_1,\lambda_1)}{\sqrt{z_g}}\,
	\mcal{H}^{lm}(z_1,\bP)\,
	\frac{k_g^m k_g^j- \delta^{mj} k_{g\perp}^2/2}
	{k_{g\perp}^2 +\mcal{M}^2}\,.\end{align}
	The functions $\varphi^{il}(z_1,\lambda_1)$ and $\mcal{H}^{lm}(z_1,\bP)$ have been previously introduced, cf. Eqs.~\eqref{phidef} and \eqref{hardamp}.
The quantity (compare to \eqn{xgdef})
\beq\label{Mdef}
	\mcal{M}^2\equiv
	z_g\left(Q^2+
	\frac{k_{1\perp}^2}{z_1} + 
	\frac{k_{2\perp}^2}{z_2} 
	\right) \simeq 
	{z_g}
	\left(Q^2 +\frac{P_{\perp}^2}{z_1 z_2} 	\right),
	\eeq
plays the role of  an effective virtuality for the gluon.

In deriving \eqn{Psig0}, we have already used the fact that the gluon is soft, $z_g\ll 1$ (and hence $z_1+z_2\simeq 1$), but at this stage $z_g$ needs not be limited by the more stringent constraint in \eqn{zg}. That is, \eqn{Psig0} also hold for the very soft gluons with $z_g\ll {\kg^2}/{P_\perp^2}$, which generate the high-energy (BK/JIMWLK) evolution. Notice that, for such very soft gluons, 
$\mcal{M}^2\ll k_{g\perp}^2$, so the virtuality plays no role. Vice-versa, in the collinear regime defined by \eqn{zg}, we have $\mcal{M}^2\gtrsim k_{g\perp}^2$, so the virtuality effects become essential. The fact of keeping $\mcal{M}^2$ within the denominator of \eqn{Psig0} amounts  to correctly treating the conservation of the LC energy  $k^-$  in the LO process $\gamma_T^* + A\to q\bar q g$. (This can be understood from the structure of \eqn{xgdef}, that can be rewritten as $ x_{q\bar q g}\simeq (\mcal{M}^2
 +\kg^2)/(z_g\hat s)$.) In turn, this will allow us to correctly reproduce the DGLAP splitting function for gluons in the target, as we shall see.

\eqn{Psig0} can be understood as follows (see \cite{Iancu:2022lcw} for details). The product
$\varphi^{il}(z_1,\lambda_1)\mcal{H}^{lm}(z_1,\bP)$ has already appeared in the LO $q\bar q$ wavefunction, \eqn{PsiLO}. As in that case, it describes the photon decay into a $q\bar q$ dipole followed by the coupling of that dipole to a low-momentum gluon --- here, the gluon {\it emitted} in the $s$-channel by either the quark, or the antiquark\footnote{Recall that in the case of \eqn{PsiLO}, the dipole-gluon coupling refers to the $t$-channel gluon exchanged with the target.}. The traceless second rank tensor with indices $(mj)$ describes the transverse momentum distribution of the emitted gluon. Its two indices account for the gluon polarisation ($j$) and for the orientation of the parent dipole in the transverse plane ($m$). It is traceless because it describes an {\it effective gluon-gluon dipole}: after emitting the gluon, the $q\bar q$ pair remains in a colour octet state whose transverse size $r\sim 1/P_\perp$ is much smaller than the transverse separation $R\equiv|\bz-\bb| \sim 1/\KT$ between this pair and the gluon. Hence, when probed on relatively large transverse separations $\gtrsim 1/R$, the three parton system indeed appears as a colour dipole made with two gluons: the physical gluon at $\bz$ and the $q\bar q$ pair at $\bb$.

This $gg$ dipole structure is also visible at the level of the scattering. A priori, the $S$-matrix involves three Wilson lines: $V_{\bx}$ for the quark, $V_{\by}^\dagger$ for the antiquark, and $U_{\bz}$ for the gluon.  (The latter is defined as in \eqn{wilson}, but with $t^a\to T^a$ for the adjoint representation.) This yields the following colour structure
\begin{align}
\label{Uin}
	U^{ac}_{\bm{z}}V_{\bm{x}}t^{c}V^{\dagger}_{\bm{y}}-t^a&\,\simeq\,
U^{ac}_{\bm{z}}V_{\bb}t^{c}V^{\dagger}_{\bb}-t^a\,=\,
(U_{\bm{z}} U^\dagger_{\bb})^{ac}t^c-t^a,
\end{align}
where we have used the fact that, on the (poor) transverse resolution scale $\sim 1/Q_s$ of the target, one can replace $V_{\bx}\simeq V_{\bb}$ and  $V^{\dagger}_{\bm{y}}\simeq V^{\dagger}_{\bb}$. As anticipated, the last expression in \eqn{Uin} involves the $S$--matrix $U_{\bm{z}} U^\dagger_{\bb}$ for a $gg$ dipole.

 At this level, one must also take into account the fact that the size $R\equiv|\bz-\bb| \sim 1/\KT$ of this dipole, albeit much larger than $r\sim 1/P_\perp$, is still much smaller than the target correlation length $ 1/Q_s$. Hence, in the leading twist approximation,  one can further expand the r.h.s. of \eqn{Uin} to linear order in $\bR$, similarly to 
\eqn{Vqq}:
\begin{align}
\label{Ugg}
(U_{\bm{z}} U^\dagger_{\bb})^{ac}t^c-t^a
\,\simeq\,-R^n\big(U_{\bz}\partial^n U_{\bz}^\dagger\big)^{ac} t^c\,.
\end{align}
Importantly, this expansion produces the same colour operator as at LO, namely $U_{\bz}\partial^n U_{\bz}^\dagger$. Hence, after the gluon emission, the WW gluon TMD appears again in the cross-section, but at the lower scale $\ell_\perp\ll \KT$.   We can recognise the premises of a renormalisation group evolution, which turns out to be DGLAP.

In order to include the effects of the collision, one must first rewrite the ``bare'' amplitude \eqn{Psig0} in transverse coordinate space. As clear from \eqn{Ugg}, which is independent of the small size $\br$, the scattering cannot modify the dependence upon the amplitude of the large relative momentum $\bP$. So, it suffices to consider  the following Fourier representation,
\begin{align}
	 (2\pi)^2\delta^{(2)}(\bK +\bm{k}_{g\perp})\,\frac{k_3^m k_3^j- \delta^{mj} k_{3\perp}^2/2}
	{k_{3\perp}^2 +\mcal{M}^2}\,=\frac{1}{2 \pi}\int \rmd^2 \bm{R}\, \rmd^2 \bm{z}\,
	e^{i \bK \cdot \bm{R} -i(\bK+\bk_{g\perp}S)\cdot \bz}
	\left(
	\frac{\delta^{mj}}{2} - \frac{R^m R^j}{R^2}
	\right)\mcal{M}^2 K_2(\mcal{M}R),
	\end{align}
and insert the scattering operator \eqn{Ugg} within the double integral in the r.h.s. This finally yields
\begin{align}\label{PsigB}
	\Psi^{ija}_{B}(\lambda_1, z_1, z_g,\bP, \bK,\ellt)	\, = i \,\frac{e e_f g  q^+}{(2\pi)^2}
		\,\frac{\varphi^{il}(z_1,\lambda_1)}{\sqrt{z_g}}\,
	\mcal{H}^{lm}( z_1, \bP)\,\mcal{G}^{mjn}(\bK,\mcal{M})	\int \rmd^2\bz\,e^{-i \ellt\cdot \bz} 
	\big(U_{\bz}\partial^n U_{\bz}^\dagger\big)^{ac} t^c,
	\end{align}
where the subscript $B$ stays for ``before (the collision)''. \eqn{PsigB} involves
the following third rank tensor
\begin{align}
	\label{Gjmn}
\mcal{G}^{mjn}(\bK,\mcal{M})
	&\,\equiv
	i\int \frac{\rmd^2 \bm{R}}{2 \pi}\,
	e^{i \bK \cdot \bm{R}}
	\left(
	\frac{\delta^{mj}}{2} - \frac{R^m R^j}{R^2}
	\right)R^n \mcal{M}^2 K_2(\mcal{M}R)\nn*[0.2cm] &\, \,=
\,\frac{\del}{\del K^n}
	\,\frac{K^mK^j- \delta^{mj} K_\perp^2/2}
	{K_{\perp}^2 +\mcal{M}^2} 
	\nn*[0.2cm]
&\, \,=\,\frac{1}{K_{\perp}^2 +\mcal{M}^2}\left[
	(\delta^{mn}K^j+\delta^{jn}K^m-\delta^{mj}K^n)-\frac{(2K^mK^j-K_{\perp}^2\delta^{mj}) K^n}{K_{\perp}^2 +\mcal{M}^2}
	\right].
 \end{align}

   \eqn{PsigB} is suggestive of TMD factorisation, but this factorisation is not yet complete: the dependences upon the transverse momenta $\bP$ and $\bK$ do not factorise from each other, due to the presence of the virtuality scale $\mcal{M}$ in \eqn{Gjmn}, which is a function of $P_\perp$, cf. \eqn{Mdef}. As we shall later argue, genuine factorisation can be recovered after re-interpreting the soft gluon as a component of the  target wavefunction.  But before doing that, we also need to compute the effect of gluon emissions in the final state (i.e. after the scattering with the SW).
      
 \subsubsection{Gluon emission after the collision}
 
 Consider now gluon emissions which occur in the final state, as shown in Fig.~\ref{fig:3jets_A}. 
 The respective contributions
 can be deduced from the general results in \cite{Iancu:2022gpw} by taking the small $z_g$ limit, while respecting the constraints in \eqn{zg}. Before we present these results, let us anticipate their physical interpretation.
 
 The scattering proceeds like in the LO process discussed in Sect.~\ref{sec:LO}: the small $q\bar q$ dipole with transverse size $r\sim 1/P_\perp$ linearly couples to the WW field $\mcal{A}^i(\bb)$ of the target and thus changes its global colour state from a singlet to an octet (while also acquiring a small transverse momentum imbalance $\ellt$). After the scattering, this small pair acts as an effective gluon which emits a (real) gluon with a much larger momentum  $\kg\gg \lt$, whose recoil fixes the momentum imbalance of the final dijet: $K_\perp\sim \kg$.
 
 The formation of the $q\bar q$ pair being described by the LO wavefunction \eqn{psiqqnew}, let us focus on the collision and the subsequent gluon emission. This is most conveniently expressed in the coordinate representation and is encoded in the following expression
  \begin{align}\label{FSG}
	\frac{ig}{(2\pi)^2}\,\frac{1}{\sqrt{z_g}}
	\left\{ \frac{x^j-z^j}{(\bx-\bz)^2}\,t^a \big[V_{\bm{x}'}V^{\dagger}_{\bm{y}} -1\big]
	- \frac{y^j-z^j}{(\by-\bz)^2}\,\big[V_{\bm{x}}V^{\dagger}_{\bm{y}'} -1\big]t^a\right\},
\end{align}
where the first term represents an emission by the quark and the second one, by the antiquark. As before,
the variables $\bx,\,\by$ and $\bz$ refer to the final coordinates of the three partons. \eqn{FSG} also involves the quark position at the scattering time $\bm{x}'$, i.e. prior to the gluon emission, which is different from $\bx$ due to the gluon recoil (and similarly for the emission by the antiquark). Namely,  $\bm{x}'$ is center of mass of the $qg$
pair produced by the splitting:
\beq
\bx'=\frac{z_1\bx+z_g\bz}{z_1+z_g}\,\Longrightarrow\,
\bx'-\bx=\frac{z_g(\bz-\bx)}{z_1+z_g}\,\simeq\,\frac{z_g}{z_1}\bR\,\Longrightarrow\,\frac{|\bx'-\bx|}{r}
\,\simeq\,\frac{z_g}{z_1}\,\frac{R}{r}\,\ll\,1,
\eeq
where we have also used $\bR=\bz-\bb\simeq \bz-\bx$, $R\sim 1/K_\perp$, $r\sim 1/P_\perp$, together with the upper limit $z_g \ll K_\perp/P_\perp$ from \eqn{zg}. To summarise, $|\bx'-\bx|\ll r$, meaning that the size
$|\bx'-\by|$ of the $q\bar q$ dipole at the time of scattering can be identified with its final size $r$ in the approximations of interest. A similar conclusion applies to $|\bx-\by'|$ in the case of the emission by the antiquark.
We can therefore simplify the expression within the accolades as
\beq\label{FSG2}
 - r^m\left\{ \frac{x^j-z^j}{(\bx-\bz)^2}\,t^a \,V_{\bb}\partial^i V_{\bb}^\dagger-
\frac{y^j-z^j}{(\by-\bz)^2}\,V_{\bb}\partial^i V_{\bb}^\dagger\,t^a\right\}\simeq
 r^m\,\frac{b^j-z^j}{(\bb-\bz)^2}\, \left[V_{\bb}\partial^i V_{\bb}^\dagger, t^a\right]
\simeq -\frac{ r^m R^j}{R^2}\, \left[V_{\bz}\partial^i V_{\bz}^\dagger, t^a\right],
\eeq
where the errors introduced by the successive approximations are of order $r/R\sim K_\perp/P_\perp$ and of order $RQ_s\sim Q_s/K_\perp$. The final result shows that, to the accuracy of interest, the dependences upon $\br$, $\bR$ and $\bz$ factorise from each other. The factor $ r^m$ contributes 
(via the Fourier transform) to the tensor  $\mcal{H}^{lm}(z_1, \bP)$, cf. \eqn{hardamp}, whereas the factor
$R^j/R^2$ yields the usual vertex for soft gluon emission in momentum space:
\beq 
\frac{i}{2\pi}\int\dif^2\bR \,e^{i\bR\cdot\bK}\,\frac{R^j}{R^2}\,=\,\frac{K^j}{K_\perp^2}.\eeq
After also using the Fierz identity $V^\dagger t^a V = U_{ac}t^c$ to rewrite the colour commutator in 
\eqn{FSG2} as
 \beq
  \left[V_{\bz}\partial^i V_{\bz}^\dagger, t^a\right] = -
\big(U_{\bz}\partial^i U_{\bz}^\dagger\big)^{ac} t^c\,,\eeq 
we end up with the following contribution to the amplitude for gluon emission {\it after} the collision:
\begin{align}\label{PsigA}
	\Psi^{ija}_{A}(\lambda_1, z_1, z_g,\bP, \bK,\ellt)	\, = -i \,\frac{e e_f g  q^+}{(2\pi)^2}
		\,\frac{\varphi^{il}(z_1,\lambda_1)}{\sqrt{z_g}}\,
	\mcal{H}^{lm}( z_1, \bP)\, \frac{K^j}{K_\perp^2}
	\int \rmd^2\bz\,e^{-i \ellt\cdot \bz} 
	\big(U_{\bz}\partial^m U_{\bz}^\dagger\big)^{ac} t^c.
	\end{align}
  
  \subsection{The target perspective: the emergence of TMD factorisation}
  \label{sec:NLOTMD}
  
  The partial results for the $q\bar q g$ LCWF, as obtained in the previous subsection, can be summarised in the following expression
  \begin{align}\label{Psigtot}
	\Psi^{ija}_{\lambda_1}(z_1, z_g,\bP, \bK,\ellt)	 = &i \,\frac{e e_f g  q^+}{(2\pi)^2}
		\,\frac{\varphi^{il}(z_1,\lambda_1)}{\sqrt{z_g}}\,
	\mcal{H}^{lm}( z_1, \bP)\left\{\mcal{G}^{mjn}(\bK,\mcal{M})-\delta^{mn}
	 \frac{K^j}{K_\perp^2}\right\}	\nonumber\\
     &\times \int \rmd^2\bz\,e^{-i \ellt\cdot \bz} 
	\big(U_{\bz}\partial^n U_{\bz}^\dagger\big)^{ac} t^c,
	\end{align}
  which is valid for $z_g$ values within the range \eqn{zg} and to ``leading-twist'', i.e. to leading order in a simultaneous expansion in the ratios $ K_\perp/P_\perp$ and $\lt/K_\perp$, assumed to be small.
  We recall that there is no {\it a priori} hierarchy between the longitudinal momentum fractions $z_1$ and $z_2$ of the quark and the antiquark, nor between the dijet relative momentum $P_\perp$ and the photon virtuality $Q$. The generic situation is such that $z_1\sim z_2 \sim 1/2$ and $P_\perp^2\sim Q^2$.
  
  One remarkable feature about \eqn{Psigtot} is the fact that the scattering between the 3-parton system ($q\bar q g$) and the target proceeds via its coupling to the WW field $\mcal{A}^n(\bz)\equiv
(i/g)U_{\bz}\partial^n U_{\bz}^\dagger\equiv \mcal{A}^i_a(\bz) T^a$ (now written in the adjoint representation), like at leading order (compare to \eqn{PsiLO}). Hence, the fully exclusive cross-section in which one measures the three ``jets'' in the finally state (that is, one measures the three transverse momenta $\bP$, $\bK$ and $\ellt$) involves again the WW TMD~\cite{Altinoluk:2020qet}, cf. \eqn{dGUU} (rewritten in terms of Wilson lines in the adjoint representation):
   \begin{align}
 \hspace*{-0.2cm}
 	\label{dGUU} \mcal{W}^{nr}(x, \ellt)
	&\ \equiv  \,\frac{1}{(2\pi)^4}
  \int\rmd^2{\bm{z}}\, \rmd^2 \overline{\bm{z}} \ 
   \rme^{-i\ellt\cdot(\bm{z}-\overline{\bm{z}})}\   \frac{-1}{\alpha_s N_c}
\left\langle \! {\rm Tr} \!\left[ 
U_{\bm{z}} (\del^n U_{\bm{z}}^{\dagger})\,
     U_{\overline{\bm{z}}} (\del^r U_{\overline{\bm{z}}}^{\dagger})\right]\right\rangle_{x}
       \end{align}
   Yet, as opposed to   \eqn{dGUU}, this is now evaluated at the softer momentum scale $\lt$ and also at a larger
   value of $x$, namely $x=  x_{q\bar q g} > x_{q\bar q}$, cf. Eqs.~\eqref{xqqdef} and \eqref{xgdef}.
   
   Here, however, we are not interested in the trijet cross-section, but rather in the ``real'' NLO correction to the dijet cross-section that is obtained from the trijet cross-section by integrating out the kinematics of the gluon at fixed values $z_1$, $\bP$ and $\bK$. That is, one must integrate over $z_g$ within the range shown in \eqn{zg} and over the soft momentum transfer $\ellt$ up to values $\sim K_\perp$. The integration over $\ellt$ generates the gluon PDF  at the scale $K_\perp$:
   \beq\label{xGK}
   xG(x, K_\perp^2) = \int\dif^2\ellt\, \Theta(K_\perp-\lt)\,\mcal{F}_g(x,\ellt)\qquad\mbox{with}\qquad
\mcal{F}_g(x,\ellt)= \mcal{W}^{nn}(x, \ellt) \,. \eeq
Also, the TMD factorisation that we would like to uncover at NLO refers to the separation of the cross-section
in the variables $\bP$ and $\bK$.  As already observed, this factorisation is not (yet) consistent with  \eqn{Psigtot}: the tensor $\mcal{G}^{mjn}(\bK,\mcal{M})$, which encodes the distribution in $\bK$, is also a function of $P_\perp$ via the quantity $\mcal{M}$ defined in \eqn{Mdef}. Clearly, this lack of factorisation at the level of the amplitude will transmit to the respective contribution to the dijet cross-section.
%
%
%

At this point, it is appropriate to recall that a TMD is, by definition, a (transverse-momentum dependent) parton distribution in the {\it target}. Hence for the $\bK$--dependence of the NLO dijet cross-section to be encoded in a TMD, the gluon previously seen as a constituent of the dipole LCWF should be reinterpreted as a {\it constituent of the target}. As recently discussed in the context of diffractive jet production \cite{Iancu:2021rup,Iancu:2022lcw,Hauksson:2024bvv}, such a reinterpretation is indeed possible for soft partons, like our gluon with $z_g\ll 1$. In practice, this amounts to a change of longitudinal variables: the ``plus'' longitudinal fraction of $z_g$, which relates the gluon emission to the incoming photon, must be replaced by its ``minus'' longitudinal fraction $x_g$ w.r.t. the target. This can be computed from the mass-shell condition
$2k_g^+k_g^-=\kg^2$, which gives
\beq\label{xg}
x_g \equiv \frac{k_g^-}{P_N^-} =\frac{\kg^2}{z_g \hat s}\ \ \Longrightarrow \ \  x_{q\bar q g} = x_{q\bar q} + x_g,
\eeq
where the last equality takes into account Eqs.~\eqref{xqqdef} and \eqref{xgdef}. This equality is suggestive of the new interpretation that we now propose (see also the second line of Fig.~\ref{fig:diagrams}): a $t$-channel (WW) gluon from the target with longitudinal momentum fraction equal to $x_{q\bar q g} $ and transverse momentum $\ellt$ splits into a gluon-gluon pair with much larger transverse momenta, $\bK$ and $\ellt-\bK\simeq -\bK$: a $t$-channel gluon with longitudinal fraction $x_{q\bar q}$, which is absorbed by the $q\bar q$ fluctuation of the photon, and an $s$-channel gluon with longitudinal fraction $x_g$, which emerges in the final state. Introducing the variable $\xi$ as the splitting fraction in the $t$-channel,
\beq\label{xidef}
\xi\equiv \frac{x_{q\bar q }}{x_{q\bar q g}}\ \ \Longrightarrow \ \ x_g = (1-\xi) x_{q\bar q g} = \frac{1-\xi}{\xi}
\,x_{q\bar q }\,.
\eeq
Comparing the two expressions for $x_g$ in Eqs.~\eqref{xg} and \eqref{xidef}, one deduces
\begin{align}
	\label{zgxi}
	z_g=\frac{\xi}{1-\xi} \frac{k_{g\perp}^2}{x_{q\bar q }\hat s}
	=\frac{\xi}{1-\xi} \frac{K_{\perp}^2}{Q^2+{P_\perp^2}/{(z_1z_2)}}\,,
\end{align} 
where in the last step we also used $\kg\simeq K_\perp$ together with \eqn{xqqdef}.  In what follows, we shall use this relation to replace $z_g$ with $\xi$ in all the previous results. In particular, it implies (cf. \eqn{Mdef})
\beq\label{Mxi}
\mcal{M}^2 = \frac{\xi}{1-\xi} K_{\perp}^2\,,\eeq
meaning that the gluon virtuality is now expressed in terms of {\it target} variables alone. Clearly, this simple change of variables achieve the separation between projectile $(z_1, \,\bP)$ and target $(\xi, \,\bK)$ variables in the expression \eqn{Psigtot} for the 3-parton amplitude, which in turn ensures a form of TMD factorisation for the cross-section. In particular, we shall replace $\mcal{G}^{mjn}(\bK,\mcal{M})\to \mcal{G}^{mjn}(\bK,\xi)$ with
\begin{align}
	\label{Gmjnxi}
\mcal{G}^{mjn}(\bK,\xi)\,=\,\frac{1-\xi}{K_\perp^2}
\left[(\delta^{mn}K^j+\delta^{jn}K^m-\delta^{mj}K^n)+(1-\xi)\left(\delta^{mj}-
\frac{2K^mK^j}{K_{\perp}^2}\right)K^n\right].
		\end{align}
To compute the cross-section, one needs to take the modulus squared of the amplitude, average over the polarisation states of the incoming photon (the index $i$), and sum over the colour  ($a$) and polarisation ($j$) states of the final gluon, as well as over the helicity ($\lambda_{1,2}$) and colour ($\alpha,\,\beta$) states of the final fermions. Let us first compute the relevant tensorial products. One has (the sum below runs over both $i$ and $\lambda=\pm 1/2$)
\beq
\varphi^{il}(z,\lambda)\,
\varphi^{i\overline{l}\,*}(z,\lambda)\,
=\,2\delta^{l\overline{l}}\left[z^2+(1-z)^2\right],\eeq
which implies that the product $\mcal{H}^{lm}   \mcal{H}^{\overline{l} \overline{m}}$   is needed only for $\bar l =l$:
\beq 
\mcal{H}^{lm}(z, P)\mcal{H}^{l\overline{m}}(z, P)=\frac{1}{(P_\perp^2 + \bar{Q}^2)^2}
	\left(\delta^{m\overline{m}} - \frac{4P^mP^{\overline{m}} \bar{Q}^2}{(P_\perp^2 + \bar{Q}^2)^2}
	\right).
	\eeq
The final tensor product that we need  is (notice that, after integrating over $\ellt$, one can restrict oneself to 	
$n=\overline{n}$)
	\begin{align}
& \left\{\mcal{G}^{mjn}(\bK,\xi)-\delta^{mn}
	 \frac{K^j}{K_\perp^2}\right\} 
\left\{\mcal{G}^{\overline{m}jn}(\bK,\xi)-\delta^{\overline{m}n}
	 \frac{K^j}{K_\perp^2}\right\}
	 \nn*[0.2cm]
&\quad \,=\,\frac{1}{K_\perp^2}\left[\delta^{m \overline{m}}
{(1-\xi)^2(1+\xi^2)} -2(1-\xi)\left(\delta^{m\overline{m}}-(1-\xi)
\frac{K^m K^{\overline{m}}}{K_\perp^2}\right)+\delta^{m \overline{m}}
\right].
	 \end{align} 
For the physical interpretation, it is useful to keep in mind that the first term inside the square brackets
arises from gluon emissions before the collision in both the direct amplitude and the complex conjugate amplitude, the last term comes from final state emissions alone, and the middle term is the interference between the two types of emissions. The terms in the last expression can be more conveniently regrouped as
\begin{align}
\frac{2}{K_\perp^2}\left[\frac{\delta^{m \overline{m}}}{2}\left(1-\xi(1-\xi)\right)^2+ (1-\xi)^2\left(\frac{K^m K^{\overline{m}}}{K_\perp^2}-\frac{\delta^{m \overline{m}}}{2}\right)\right].
 \end{align} 
 By inspection of the above results, it should be clear by now that the ensuing cross-section will bring NLO contributions to both the unpolarised and the linearly polarised gluon TMDs, with the same hard factor as at leading order (recall \eqn{sigma0}). 
 
 To compute the dijet cross-section, one also needs to integrate over the gluon longitudinal fraction
 $z_g$. After the change of variables $z_g\to \xi$,  this integration becomes (for a generic function $f(z_g)$)
 \beq\label{Jacob}
 \int_{x_*\frac{\KT^2}{\PT^2}}^{\frac{\KT}{\PT}}\frac{\rmd z_g}{z_g} \,f(z_g)=\,
 \int_{x_\star}^{1-\xi_0(K_{\perp}/P_\perp)}\frac{\rmd\xi}{\xi(1-\xi)}\,f\big((z_g(\xi)\big)\,,\qquad
 \xi_0(K_{\perp}/P_\perp)\,\equiv\,\frac{\KT}{\PT}\,,
 \eeq
 where the integration limits are obtained by combining Eqs.~\eqref{zg} and \eqref{zgxi}, and we have also used the fact that both $x_*$ and $\KT/\PT$ are small numbers. Here and in what follows, we neglect corrections suppressed by powers of $x_*$ and/or $ \xi_0$: it is only the logarithmic sensitivity  to any of these quantities that matters.
 
 At this point, we are prepared to deduce the final result for this particular correction (the leading twist part of the NLO contribution due to a real gluon emission) to the cross-section for hard dijet production.
 A straightforward calculation yields
 \begin{align}
	\label{sigmaNLOR}
	\frac{\dif \sigma^{\gamma_{T}^* A \to q\bar{q}X}_{\nlo, \mcal{R}}}
	{\rmd z_1
  	\rmd z_2\dif^2\bP\dif^2\bK } &\,=  {\alpha_{em}\alpha_s}
	e_f^2\delta(1-z_1-z_2)\left(z_1^{2} + 
	z_2^{2}\right) \frac{P_{\perp}^4 + \bar{Q}^4}
	{(P_{\perp}^2 + \bar{Q}^2)^4}\nonumber\\*[0.2cm]
	&\, \times \left\{\,\Delta \mcal{F}_{\mcal{R}}^{(1)}(x,K_{\perp}, P_\perp^2)
	 -   \frac{4 P_{\perp}^2 \bar{Q}^2}{P_{\perp}^4 + \bar{Q}^4}
        \left[ \frac{( \bP\cdot \bK)^2}{P_{\perp}^2\,K_{\perp}^2 }
       -\frac{1}{2}
        \right]  \Delta\mcal{H}_{\mcal{R}}^{(1)}(x,K_{\perp})\right\}.
\end{align}
As anticipated, this has the same structure as the LO result in \eqn{sigma0} --- in particular, it exhibits the same hard factors in both the unpolarised and the linearly polarised sector ---, but with modified expressions for the respective gluon TMDs which express ``real'' NLO corrections in the approximations of interest. 

Namely, for the  unpolarised TMD, one finds
 \begin{align}
	\label{gTMDreal}
	\Delta \mcal{F}_{\mcal{R}}^{(1)}(x, K_{\perp}, P_\perp^2) =
	 \frac{\alpha_s }{2\pi^2 K_\perp^2}\!
\int_{x_\star}^{1-\xi_0(K_{\perp}/P_\perp)}\!\!\rmd\xi \,P_{gg}\left(\xi\right)\ \frac{x}{\xi} \,G^{(0)}\left(\frac{x}{\xi}, 
K_\perp^2\right). \end{align}
As emphasised by the notations, this NLO correction also depends upon the hard momentum $P_\perp$, via the upper limit on the integral over $\xi$. The LO gluon PDF $xG^{(0)} (x, \KT^2)$ has been defined in \eqn{PDF0}.
Notice that, in general,  ``leading order'' is not synonymous of ``tree-level approximation'' (say, as provided by the MV model): the LO approximation also includes the effects of the BK/JIMWLK evolution down to the value $x$ of interest. In \eqn{gTMDreal}, this value is written as $x/\xi$, where it is understood that  $x\equiv x_{q\bar q}$;  hence the argument $x/\xi$ of the LO gluon TMD under the integral takes the physical value $x_{q\bar q g}$, cf. \eqn{xidef}.
The final ingredient in  \eqn{gTMDreal} that needs to be explained is $P_{gg}(\xi)$; this is defined as 
 \begin{align}
\label{eq:Pgg}
P_{gg}(\xi)&\equiv 2N_c\,  \frac{[1-\xi(1-\xi)]^2}{\xi(1-\xi)}
= 2N_c\,  \frac{1+ (1-\xi)^2(1+\xi^2) -(1-\xi^2)}{\xi(1-\xi)}\,,
	\end{align} 
and is recognised as the unregularised $g\to gg$ DGLAP splitting function. 
Although  obtained in the {\it projectile} picture, i.e. by computing gluon emissions by the quark and the antiquark, \eqn{gTMDreal} is naturally interpreted as the effect of a hard gluon splitting which occurs in the {\it target} wavefunction and in the $t$-channel (see the second row in Fig.\,\ref{fig:diagrams}):
a gluon with  transverse momentum $\ell_\perp\ll \KT$ splits into a pair of harder gluons with momenta $\pm\bK$.

Whereas the appearance of this splitting function (together with the characteristic spectrum $1/\KT^2$) would be natural when computing a hard splitting in the {\it target} wavefunction, its emergence from the {\it dipole} evolution is highly non-trivial. This is also illustrated by the decomposition of the numerator in the second equality in \eqn{eq:Pgg}: the three pieces in this decomposition correspond to (dipole-picture) emissions occurring in the final state, in the initial state, and to interferences between the two, respectively, see Figs.\,\ref{fig:3jets_B}-\ref{fig:3jets_A}. The denominator exhibiting the characteristic singularities at $\xi=0$ and $\xi=1$ has been generated via the Jacobian for the change of  variables from $z_g$ to $\xi$, cf. \eqn{Jacob}. It is also interesting to observe that the singular piece 
 $\propto 1/[\xi(1-\xi)]$ in $P_{gg}(\xi)$  exclusively comes from final-state emissions in the dipole picture.  The sum of initial state emissions and of the interference terms is responsible for the regular piece of $P_{gg}(\xi)$.
 
The respective correction to the linearly polarised TMD reads as follows
 \begin{align}
	\label{gHreal}
	\Delta \mcal{H}_{\mcal{R}}^{(1)}(x, K_{\perp}) =
	 \frac{\alpha_s }{2\pi^2 K_\perp^2}\!
\int_{x_\star}^{1} \rmd\xi \,\frac{2N_c(1-\xi)}{\xi}\ \frac{x}{\xi} \,G^{(0)}\left(\frac{x}{\xi}, 
K_\perp^2\right). \end{align}
This has no singularity at $\xi\to 1$, hence one can ignore the respective physical cutoff $\xi_0$, which in turn implies that $\Delta \mcal{H}_{\mcal{R}}^{(1)}$ does not depend upon the hard scale $P_\perp$. \eqn{gHreal} 
gives a  contribution to the polarised distribution $ \mcal{H}_g(x, K_{\perp})$ generated via a hard splitting of a gluon with any transverse momentum $\lt\ll \KT$. The associated splitting function turns out to be $2N_c(1-\xi)/\xi$. This is perhaps less familiar than the contribution to  $ \mcal{F}_g(x, \bK)$ in \eqn{gTMDreal}, but it was nevertheless known and can be found e.g. in Eq.~(25) of Ref.~\cite{Sun:2011iw}.

We conclude this subsection with a discussion of the role of the rapidity divider $x_*$ in equations like \eqn{gTMDreal} and \eqn{gHreal}. Recall that we consider a small-$x$ context: the variable $x\equiv x_{q\bar q}$ which appears in these equations is so small  (typically $x < 10^{-2}$) that $\alpha_s\ln(1/x)\gtrsim 1$, hence one needs to resum the radiative corrections (soft gluon emissions) to the target wavefunction which are accompanied by leading powers of $\ln(1/x)$. This can be done by solving the BK/JIMWLK equations for the unpolarised WW distribution $\mcal{F}_g^{(0)}(x,\ellt)$ down to the value of $x$ of interest (starting e.g. with an initial condition at $x_0=10^{-2}$ as taken from the MV model). Notice that the typical values of $\lt$ in the gluon distribution at $x$ is of the order of the saturation momentum $Q_s(x)$. This scale increases quite fast with decreasing $x$, but remains a semi-hard scale (in the ballpark of 1 to 2 GeV) for the values of $x$ that are relevant for the phenomenology. So, the {\it typical} momentum imbalance that a $q\bar q$ dijet would acquire via scattering off the nuclear targte is $K_\perp\sim Q_s(x)$. What if one is interested in considerably larger values $K_\perp\gg Q_s(x)$ ? They can be naturally created via one (or few) hard splittings inside the target wavefunction, which  occur at the end of the high-energy evolution (to maximise the effects of the latter). Eqs.~\eqref{gTMDreal} and \eqref{gHreal} describe one such a splitting and can be promoted to evolution equations (which resum several such splittings), as we shall argue in Sect.~\ref{sec:evol}. 

This additional evolution, of the DGLAP type, must be restricted to $x'$ values which are close enough to the final value $x$, say $x'\le x/x_*$ with $\alpha_s\ln(1/x^*)\ll 1$, to minimise the effects of the high energy evolution. Rewriting $x'=x/\xi$, with $\xi$ the splitting fraction in the $t$-channel, we deduce $\xi > x_*$, like in Eqs.~\eqref{gTMDreal} and \eqref{gHreal}.
Correspondingly, the BK/JIMWLK evolution of the LO gluon TMD $\mcal{F}_g^{(0)}(x',\ellt)$ must be restricted to the interval $x_0 > x' > x/x_*$, to avoid an overlap in $x'$ with the DGLAP evolution. As we shall later argue, these constraints are sufficient to ensure that the dijet cross-section is independent of the arbitrary divider  $x_*$ to the logarithmic accuracy of interest: the contributions enhanced by powers of $\alpha_s\ln(1/x^*)\ln(\KT^2/Q_s^2)$ cancel between the high-energy evolution of $\mcal{F}_g^{(0)}(x',\ellt)$ and the DGLAP evolution of $xG(x,\KT^2)$, to be fully uncovered later.




\subsection{Comments on the virtual corrections: the $\beta_0$ term}
\label{sec:beta}

A complete evolution also requires the corresponding virtual corrections --- those associated with gluon emissions within the range $\KT\ll \kg\ll \PT$. To the accuracy of interest, these are quite simple: besides the double-logarithmic Sudakov factor  computed in Eq.\,\eqref{FSukvirt} (as also found in the full 
NLO calculations~\cite{Caucal:2021ent,Taels:2022tza} , see e.g. Eq.~(8) in \cite{Caucal:2023fsf}), there is a single-logarithmic piece, proportional to  $\beta_0=11/12- N_f/6N_c $ (the coefficient of the one-loop $\beta$--function), which arises from the RG flow of the gluon TMD from the gluon transverse momentum $\KT^2$ up to the renormalisation scale $\mu_R^2\sim \PT^2$:
\begin{align}
	\label{gTMDvirt}
	\Delta \mcal{F}_{\mcal{V}}(x, K_\perp,\PT^2)   = 	
\frac{\alpha_s N_c }{\pi}\left[-\frac{1}{4}\ln^2\frac{\PT^2}{\KT^2}+\beta_0\ln\frac{\PT^2}{\KT^2}\right]
\mcal{F}_g^{(0)}(x, K_{\perp}). \end{align}

The $\beta_0$--piece is not naturally generated by the NLO calculation
in the dipole picture \cite{Taels:2022tza,Caucal:2023nci,Caucal:2023fsf}, because of the classical approximation used for the scattering~\cite{Xiao:2017yya,Hentschinski:2021lsh}. In the target picture, this piece is well known to arise via one-loop  corrections to the gluon exchange in the $t$-channel  \cite{Ayala:1995hx,Zhou:2018lfq,Mueller:2018llt}. (In the target light-cone $A^-=0$, one just needs to consider the self-energy insertions, like the gluon loop in Fig.~\ref{fig:diagrams} \cite{Mueller:2018llt}.)

 Without any calculation, the need for this contribution can be understood by inspection of the LO result \eqn{sigma0int} for the dijet cross-section integrated over $K_\perp$ (see also the discussion at the end of section 5.1 in \cite{Caucal:2023nci}). This result involves the product $\alpha_s \,xG(x,\PT^2)$, where we recall that $\alpha_s$ is the coupling  between the $q\bar q$ dipole and a $t$-channel gluon from the target, and $xG(x,\PT^2)$ is the gluon PDF.  
It is  {\it a priori}  clear that the NLO corrections must include the one-loop running coupling (RC) correction to $\alpha_s$ (on physical grounds, this must be evaluated at the hard scale $\PT^2$), together with one step in the renormalisation group (DGLAP) evolution of the gluon PDF. Yet, it turns out that some of these corrections are not reproduced by the standard NLO calculations within the CGC effective theory~\cite{Caucal:2021ent,Taels:2022tza}: they miss the RC corrections to $\alpha_s$ together with the virtual piece proportional to $\beta_0\delta(1-\xi)$ from the DGLAP splitting function. These two corrections are in fact  related to each other, as well known and it will be briefly reviewed now.


To that aim, it is convenient to resort again on the perspective of target evolution and consider the one-loop loop corrections to the gluon exchange in the $t$-channel,  cf. Fig.~\ref{fig:diagrams}. It is convenient to use the target light-cone gauge $A^-=0$, since in this particular gauge\footnote{In other gauges, such as the Feynman gauge, the one-loop $\beta$--function also receives contributions from the quark-gluon vertex corrections, from the quark self-energies, and from the insertion of a ghost loop in the gluon propagator.} the ultraviolet divergences associated with colour charge renormalisation are limited to the quark and gluon loop insertion in the gluon propagator \cite{Mueller:2018llt}. As standard in pQCD, we use dimensional regularisation with mass parameter $\mu$ and implement  the UV renormalisation via minimal subtraction. As a result, the leading-order  gluon propagator with bare gauge coupling $\alpha_{s0} = g^2_0/4\pi$ (one power of $g_0$ at each end of the propagator) gets replaced by the following renormalised result,
   \begin{align}\label{propnlo}
	\frac{\alpha_{s0} }{\kg^2}
	\to
	\frac{\alpha_\mu}{\kg^2}
	\left[1 - \beta_0\frac{\alpha_\mu N_c}{\pi}  \ln\frac{\kg^2}{\mu^2}\right]\,.
	 \end{align}
where  $\alpha_\mu\equiv \alpha_s(\mu^2)$ is the one-loop RC at the renormalisation scale $\mu$, defined as
 \begin{align}
	\label{1LRC}
\alpha_s(\mu^2)\,=\, 
\frac{\pi}{\beta_0 N_c\ln({\mu^2}/{\Lambda^2_{\rm QCD}})}
	 \qquad \mathrm{with} \qquad
	 \beta_0 \equiv \frac{11}{12} -\frac{N_f}{6N_c}
	  \,.
\end{align}
When computing a cross-section at NLO, one must include the one loop corrections in either the direct amplitude, or the complex conjugate amplitude. Then the r.h.s. of \eqn{propnlo} gets replaced by
\begin{align}\label{Xsecnlo}
		\frac{\alpha_\mu^2}{\kg^4}
	\left[1 - 2\beta_0\frac{\alpha_\mu N_c}{\pi}  \ln\frac{\kg^2}{\mu^2}\right]\,.
	 \end{align}
One factor of $\alpha_\mu$ is associated with the gluon coupling to the projectile (the $q\bar q$ dipole) and the other one with its coupling to a parton from the target (a valence quark, or the parent parton in one step of the quantum evolution). The virtual corrections proportional to $\beta_0$ must somehow renormalise both couplings, but the proper way to distribute them turns out to be quite subtle. Namely, it is quite natural to use half of these corrections to replace the coupling to a parton in the target by its value at the scale $\kg^2$:
\beq
\alpha_\mu\,\to \,\alpha_s(\kg^2)\,=\,\alpha_\mu\left[1 - \beta_0\frac{\alpha_\mu N_c}{\pi}  \ln\frac{\kg^2}{\mu^2}\right].\eeq
Indeed, the parent parton typically has a much lower transverse momentum $\lt\ll\kg$, hence the hard scale that should control the parton-gluon coupling  is clearly $\kg$. On the other, the upper end of the gluon propagator is attached to a fermion with much larger momentum $\PT\gg \kg$, so the respective RC should rather be evaluated at this hard scale $\PT^2$. To that aim, we shall rewrite the remaining half of the virtual correction in \eqn{Xsecnlo} as
\beq
\beta_0\frac{\alpha_\mu N_c}{\pi}  \ln\frac{\kg^2}{\mu^2}= \beta_0\frac{\alpha_\mu N_c}{\pi}  \ln\frac{\PT^2}{\mu^2}
-\beta_0\frac{\alpha_\mu N_c}{\pi}  \ln\frac{\PT^2}{\kg^2}\,.\eeq
The first term in the r.h.s. is now used to replace the coupling to the hard quark as follows 
\beq
\alpha_\mu\,\to \,\alpha_s(\PT^2)\,=\,\alpha_\mu\left[1 - \beta_0\frac{\alpha_\mu N_c}{\pi}  \ln\frac{\PT^2}{\mu^2}\right].\eeq
Hence, to the order of interest, the product of couplings in \eqn{Xsecnlo} has been rewritten as
\begin{align}\label{Xsecnlo2}
		\alpha_\mu^2
	\left[1 - 2\beta_0\frac{\alpha_\mu N_c}{\pi}  \ln\frac{\kg^2}{\mu^2}\right]\simeq
	\,\alpha_s(\kg^2)\alpha_s(\PT^2)\left[1 + \beta_0\frac{\alpha_\mu N_c}{\pi}  \ln\frac{\PT^2}{\kg^2}\right].
	 \end{align}
The remaining virtual correction as shown in the above square brackets is a contribution to the DGLAP evolution
of the quark TMD from the gluon scale $\kg^2$ up to the dijet scale $\PT^2$. This is recognised as the Sudakov single log in Eq.~\eqref{gTMDvirt}.  After integrating over $\kg^2$ up
to $\PT^2$, this generates the expected virtual contribution  $\beta_0\delta(1-\xi)$ to the DGLAP splitting function, as we will verify in the next section. 

The above procedure demonstrates the consistency between collinear factorisation and running coupling corrections at one-loop order. So, clearly, the inclusion of the one-loop corrections proportional to $\beta_0$ is compulsory in order to obtain a consistent result at NLO.  At this point, one may wonder why these corrections are not automatically generated by the standard perturbative calculations in the CGC approach. 
As shown by the above manipulations, based on the target picture, the corrections associated with charge renormalisation refer to the propagator of the exchanged gluon. But in the CGC approach, this gluon is treated as a classical background field (the ``shockwave''). By itself, this is not a limitation of principle: one should be able to renormalise this classical field and include the relevant RC corrections. A similar renormalisation was needed for the purposes of the high-energy evolution (the ``sigma'' term in the construction of the JIMWLK equation
\cite{JalilianMarian:1997jx,JalilianMarian:1997gr,Kovner:2000pt,Iancu:2000hn,Iancu:2001ad,Ferreiro:2001qy}).
One should be able to extent this procedure beyond the leading logarithmic approximation at small $x$. This has not been done yet and it goes beyond the scope of this work. Rather, we shall simply add this correction to our results, thus following a strategy that was also adopted in other recent papers~\cite{Caucal:2023nci,Caucal:2023fsf}.


\section{From  NLO corrections to the DGLAP and CSS evolutions}
\label{sec:evol}

The (real and virtual) NLO corrections computed in the previous section can be summarised in the following approximation for the unpolarised gluon WW TMD, valid in the leading twist approximation at sufficiently large transverse momenta  $P_\perp \gg K_\perp \gg Q_s(x)$:
 \beq\label{FNLO}
\mcal{F}_g^{(1)} (x, K_{\perp}, P_\perp^2) =\mcal{F}_g^{(0)}(x/x_*, K_{\perp}) + \Delta \mcal{F}_{\mcal{R}}^{(1)}(x, K_{\perp}, P_\perp^2) +\Delta \mcal{F}_{\mcal{V}}^{(1)}(x, K_{\perp}, P_\perp^2) \,,\eeq
where the argument $x/x_*$ of the LO contribution reflects the rapidity phase-space for the high energy evolution, cf. \eqn{smallx}. We recall that the dependence upon the hard momentum scale $P_\perp$ (the relative momentum of the $q\bar q$ jets) enters via the one-loop corrections, and more precisely via the limits \eqn{zg} on the longitudinal phase-space for the soft gluon and also (in the case of the virtual corrections) via the upper limit $P_\perp^2$ on the integral over the gluon transverse momentum $\kg^2$. Let us also specify the definition of the  integrated distribution  (the gluon PDF) in the presence of this additional hard-scale dependence; this reads
\begin{align}\label{PDF}
xG(x, \PT^2) \,=\pi \int^{\PT^2}_{\Lambda^2} {\rmd K_\perp^2}\,\mcal{F}_g(x, K_{\perp}, \PT^2).
 \end{align}
 Importantly, the variable $P_\perp$ enters not only the upper limit, but also the argument of the gluon TMD inside the integrand.
 
We shall now argue that the $P_\perp$-dependence of the one-loop corrections can be recognised as one step in the combined DGLAP+CSS evolutions. As usual, the DGLAP evolution~\cite{Gribov:1972ri,Altarelli:1977zs,Dokshitzer:1977sg}  refers to the gluon PDF, while the CSS evolution (for which our approach provides only a special limit) refers to the gluon TMD~\cite{Collins:1981uk,Collins:1981uw,Collins:1984kg,Collins:2011zzd,Boussarie:2023izj}.

\subsection{The gluon PDF: the emergence of the DGLAP evolution}
\label{sec:DGLAP}

The first step in that sense is to explicitly extract the $P_\perp$-dependence of the real correction $\Delta \mcal{F}_{\mcal{R}}^{(1)}$, which enters via the upper limit of the integral over $\xi$ in \eqn{gTMDreal}. This can be easily done with the help of the plus prescription, defined as (for a generic function $f(\xi)$ which is finite at $\xi=1$)
\beq
\int_{x_\star}^{1}\rmd\xi\,\frac{f(\xi)}{\big(1-\xi \big)_+}=\int_0^1
\rmd\xi\,\frac{\theta(\xi-x_\star)f(\xi)-f(1)}{1-\xi}\,=\lim_{\xi_0\to 0}
\left\{\int_{x_\star}^{1-\xi_0}\rmd\xi\,\frac{f(\xi)}{1-\xi}
-f(1)\int_{0}^{1-\xi_0}\frac{\rmd\xi}{1-\xi}\right\}.
\eeq
This implies 
\beq
\int_{x_\star}^{1-\xi_0}\rmd\xi\,\frac{f(\xi)}{1-\xi}=\int_{x_\star}^{1}\rmd\xi\,\frac{f(\xi)}{\big(1-\xi \big)_+}
+f(1)\ln\frac{1}{\xi_0}+\order{\xi_0}\,,\eeq
which can be now applied to rewrite the r.h.s. of  \eqn{gTMDreal} as (up to corrections suppressed by powers
of $K_\perp/P_\perp$)
\begin{align}
	\label{eq:Dreal}
	 \Delta \mcal{F}_{\mcal{R}}^{(1)}(x, K_{\perp}, P_\perp^2) = 	
	 \frac{\alpha_s }{2\pi^2 K_\perp^2}
\int_{x_\star}^{1}\rmd\xi\,{P}^{(+)}_{gg}(\xi)\,\frac{x}{\xi}G^{(0)}\left(\frac{x}{\xi}, K_\perp^2\right)+
\frac{\alpha_s N_c}{2\pi^2 K_\perp^2}\ln\frac{\PT^2}{K_{\perp}^2}\,xG^{(0)}(x, \KT^2).
 \end{align}
Here ${P}^{(+)}_{gg}$ differs from  \eqn{eq:Pgg} only in the replacement  $(1-\xi)\to (1-\xi)_+$ in the denominator:
\beq\label{Pggreg}
P_{gg}^{(+)}\left(\xi\right)=2N_c\left\{\frac{\xi}{\big(1-\xi \big)_+}+\frac{1-\xi}{\xi} +\xi(1-\xi)\right\}.
\eeq
The second tern in \eqn{eq:Dreal} is recognised as the real Sudakov contribution in Eq.\,\eqref{FSukreal}.

Let us also recall here the corresponding virtual contribution, cf. Eq.~\eqref{FSukvirt} and Eq.~\eqref{gTMDvirt} which includes the $\beta_0$ single logarithm:
\begin{align}
	\label{gTMDV}
	\Delta \mcal{F}_{\mcal{V}}^{(1)}(x, K_{\perp}, P_\perp^2)= -\frac{\alpha_s N_c }{\pi}\left(\frac{1}{4}\ln^2\frac{\PT^2}{\KT^2}-\beta_0\ln\frac{\PT^2}{\KT^2}\right)
\mcal{F}_g^{(0)}(x, K_{\perp}). \end{align}
Strictly speaking, this result is only valid for $K_\perp\gg Q_s$, as recalled at the beginning of this section. As shown in~\cite{Mueller:2013wwa,Taels:2022tza,Caucal:2023fsf}, the Sudakov logarithms in the saturation regime appear at NLO in coordinate space, i.e.~$-\alpha_s N_c/(4\pi) \ln^2((P_\perp^2(\bt-\bt')^2)/c_0^2)$ with $c_0=2e^{-\gamma_E}$ for the double logarithm. It is only in the regime $K_\perp \gg Q_s$ that one can replace $(\bt-\bt')^2/c_0^2\to 1/K_\perp^2$ as discussed in subsection~\ref{sub:TMD-cs}. Note also that in addition to the terms shown in Eq.\,\eqref{gTMDV}, there are other single logarithms (and pure $\alpha_s$ corrections without logarithmic enhancement) which are not associated with the DGLAP+CSS evolution of the WW gluon TMD, see Eq.\,(11) in \cite{Caucal:2023fsf}.

Using the above, one can deduce the one-loop correction to the gluon PDF Eq.\,\eqref{PDF} and thus recognise the first step in the DGLAP evolution. To that aim, one must integrate \eqn{FNLO} over $K_\perp^2$ up to $\PT^2$. The key observation is that, after this integration, the effect of the real Sudakov in \eqn{eq:Dreal} cancels against that of the double logarithm in \eqn{gTMDV}. Indeed, let us start with the latter:
\begin{align}\label{rate0}
-\frac{\alpha_s N_c }{4\pi}\int^{\PT^2}_{\Lambda^2} {\rmd K_\perp^2}\,\ln^2\frac{\PT^2}{\KT^2}\,\mcal{F}_g^{(0)}(x, K_{\perp})&=
-\frac{\alpha_s N_c }{2\pi}\int^{\PT^2}_{\Lambda^2} {\rmd K_\perp^2}\,\mcal{F}_g^{(0)}(x, K_{\perp})
\int_{\KT^2}^{\PT^2}\frac{\dif\lt^2}{\lt^2}\,\ln\frac{\PT^2}{\lt^2}\nn*[.2cm]
&=-\frac{\alpha_s N_c }{2\pi}\int^{\PT^2}_{\Lambda^2} \frac{\dif\lt^2}{\lt^2}\,\ln\frac{\PT^2}{\lt^2}
\int^{\lt^2}_{\Lambda^2} {\rmd K_\perp^2}\,\mcal{F}_g^{(0)}(x, K_{\perp})
\nn*[.2cm]
&=-\frac{\alpha_s N_c }{2\pi^2}\int^{\PT^2}_{\Lambda^2} \frac{\dif\lt^2}{\lt^2}\,\ln\frac{\PT^2}{\lt^2}\,
xG^{(0)}(x,\lt^2)\,,
\end{align}
where in going from the first to the second line, we have reversed the order of the integrations. As anticipated, the final result is the same as minus the integral of the real Sudakov in \eqn{eq:Dreal}. This precise compensation should not be seen as a surprise, but rather as a consistency check for our calculation. Recall that, in the construction of the DGLAP equation, soft and collinear divergences are supposed to cancel between real and virtual corrections. This is indeed the meaning of the above  cancellation, which merely demonstrates the mutual consistency of the physical cutoffs that we have chosen for the real and the virtual corrections, respectively. This consistency was already clear from our unified treatment of the two types of corrections within the dipole picture, cf. Eqs.~\eqref{Frealfin}-\eqref{FSukreal}.
After this cancellation, the contribution of  \eqn{FNLO} to the gluon PDF is obtained as
\begin{align}\label{PDFnlo}
xG^{(1)}(x, \PT^2) \,=\,xG^{(0)}(x/x_\star, \PT^2)+ \frac{\alpha_s }{2\pi^2} \int^{\PT^2}_{\Lambda^2} 
\frac{\rmd K_\perp^2}{K_\perp^2}
\int_{x_\star}^{1}\rmd\xi\,\mcal{P}_{gg}\,\frac{x}{\xi}G^{(0)}\left(\frac{x}{\xi}, K_\perp^2\right),
 \end{align}
with $\mcal{P}_{gg}$ the full (regularised) DGLAP splitting function  \cite{Collins:2011zzd}, that is
\beq\label{Pggreg}
\mcal{P}_{gg}\left(\xi\right)=2N_c\left\{\frac{\xi}{\big(1-\xi \big)_+}+\frac{1-\xi}{\xi} +\xi(1-\xi)+\beta_0\delta(1-\xi)\right\},
\eeq
where the piece proportional to $\beta_0$ has been generated by the respective virtual correction in 
\eqn{gTMDV}. 

\eqn{PDFnlo} is recognised as one step in the DGLAP evolution of the gluon PDF, an anticipated.
Clearly, its generalisation beyond one-loop order is the DGLAP equation, that we write in differential form:
 \begin{align}\label{DGLAP}
 \frac{\del xG  (x, \PT^2)}{\del \ln \PT^2}\,=\,\pi \PT^2 \mcal{F}^{(0)}\left(\frac{x}{x_*}, \PT\right)+
 \frac{\alpha_s(\PT^2) }{2\pi}
\int_{x_\star}^{1}\rmd\xi\,\mcal{P}_{gg}\left(\xi\right)
\,\frac{x}{\xi}
G\left(\frac{x}{\xi}, \PT^2\right).
\end{align}
We have also inserted the argument of the running coupling, since this is well known in this particular case. 

By inspection of  \eqn{DGLAP}, one can notice two important differences w.r.t. the standard version of the DGLAP equation in the literature~\cite{Gribov:1972ri,Altarelli:1977zs,Dokshitzer:1977sg} :

\texttt{(i)} the presence of a {\it source term} in the r.h.s., namely the LO gluon WW TMD $ \mcal{F}^{(0)}(x/x_*, \PT^2)$, that is computed in the standard CGC approach, including the B/JIMWLK evolution  
from $x_0\sim 10^{-2}$ (where one needs a model for the initial condition, such as the MV model~\cite{McLerran:1993ni,McLerran:1993ka}) down to 
the interesting value $x\ll x_0$ (with $x=x_{q\bar q}$ for the djet problem at hand);

\texttt{(ii)} the lower limit $x_*$ in the integral over $\xi$ instead of the standard value $\xi_{\rm min}=x$;
clearly, we are in a small-$x$ set-up where $x_*\gg x$, so the longitudinal phase-space available to this evolution is conserably reduced by this constraint. As previously explained, this is indeed needed in order to avoid the overlap with the high-energy (B/JIMWLK) evolution of the target gluon distribution.

As it should be clear from the above discussion, \eqn{DGLAP} does not describe the standard DGLAP evolution at moderate $x$, but rather a special kind of DGLAP evolution which arises on top of the high-energy evolution in small-$x$ problems where the observables carry relatively large transverse momenta  $\PT\gg Q_s(x)$. The BK/JIMWLK equation occupies most of the rapidity interval $Y\sim \ln(1/x)\gg 1$ between  the projectile and the target, whereas the additional, DGLAP-like, evolution is restricted to a small interval $\Delta Y\sim \ln(1/x_*)\sim \order{1}$ towards the rapidity of the projectile.

The formulation of the DGLAP evolution  as a source problem  is natural in this small-$x$ context, where the CGC calculations predict a ``tree-level'' value for the unintegrated gluon distribution $\mcal{F}^{(0)}(x, \bP)$. A similar formulation has been proposed  in \cite{Iancu:2022lcw,Iancu:2023lel,Hauksson:2024bvv} in the context of DIS diffraction. Clearly, this 
evolution should become effective only for sufficiently large transverse momenta $\PT\gg Q_s$, where the source term $\mcal{F}^{(0)}(x/x_*, \PT)$ develops the perturbative power-tail $\propto 1/\PT^2$. Accordingly, the initial condition for \eqn{DGLAP} must be formulated at some scale $\mu_0$ which is comparable to, but larger than, the saturation momentum, e.g. $\mu_0=2 Q_s(x)$. In practice one should vary this scale and interpret the variance of the solution as a form of scheme dependence, that should decrease when increasing $\PT^2/\mu_0^2$. Given the structure of \eqn{DGLAP}, it is clear that the appropriate initial condition reads (recall \eqn{PDF})
\beq\label{ICdglap}
xG(x, \mu_0^2) = xG^{(0)}(x, \mu_0^2) \equiv \pi \int^{\mu_0^2}_{\Lambda^2}\!{\rmd \ell_\perp^2}
 \mcal{F}^{(0)}(x, \lt).\eeq
As clear from the r.h.s. of \eqn{DGLAP}, this initial condition is needed for values of $x$ within the range $x_{q\bar q}< x < x_{q\bar q}/x_*$. Accordingly, the  scheme dependence associated with the value $\mu_0$ is of the order $\alpha_s\ln(1/x_*)\ln(\mu_0^2/Q_s^2)$, which is parametrically small so long as $\mu_0\sim Q_s$.

 Hence, for $\PT^2 <\mu_0^2$, the gluon PDF  $xG  (x, \PT^2)$ reduces to the LO result $xG ^{(0)} (x, \PT^2)$ given by the CGC approach, while for larger momenta $\PT^2 > \mu_0^2$, it receives contributions from both the source term and the DGLAP evolution. The source term is essential at small $x$, since it describes the gluon multiplication via the high-energy evolution. But for sufficiently large $\PT\gg Q_s(x)$, the DGLAP evolution becomes important as well and its results are sensitive to the small-$x$ dynamics (in particular, to gluon saturation) via the initial condition \eqn{ICdglap}.

We conclude this section with a discussion of the sensitivity of the DGLAP equation \eqn{DGLAP} to the arbitrary rapidity divider $x_*$.  We would like to argue that its solution $ xG(x, \PT^2) $ does not include potentially large contributions  enhanced by powers of $ \alpha_s  \ln(1/x_*) \ln(\PT^2/Q_s^2)$.  Such contributions would be the only ones to be harmful (i.e.~to introduce a non-trivial sensitivity to $x_*$) in the approximations of interest. The argument becomes more transparent if one first considers 
the evolution equation in the double-logarithmic approximation (DLA)  --- a common limit of BK and DGLAP evolutions valid at small $x$ and large $\PT\gg Q_s(x)$. This reads
\begin{align}\label{DLA}
 \frac{\del  xG(x, \PT^2) }{\del \ln \PT^2\del \ln(1/x)}=\frac{\alpha_s(\PT^2) N_c}{\pi}\,xG(x, \PT^2) \,.
\end{align} 
This should be integrated over $x$ from $x_0$ (the largest value of $x$ at which one fixes the initial condition) down to interesting value $x\ll 1$. One can separate this integration into two intervals, from $x_0$ down to $x_1$, and from $x_1$ down to $x$, with $x_1$  an arbitrary value between $x$ and $x_0$. The overall result of such a piecewise evolution is of course independent of $x_1$. We can successively write
 \begin{align}\label{DLA1}
 \frac{\del  xG(x, \PT^2) }{\del \ln \PT^2}&\,=\frac{\alpha_s(\PT^2) N_c}{\pi}\,\int_{x}^{x_0}
 \frac{\rmd x'}{x'}\,x'G(x', \PT^2) \nn*[0.2cm]
 &\,=\frac{\del  x_1G(x_1, \PT^2) }{\del \ln \PT^2} \,+\,\frac{\alpha_s(\PT^2) N_c}{\pi}\,\int_{x}^{x_1}
 \frac{\rmd x'}{x'}\,x'G(x', \PT^2) 
 \nn*[0.2cm]
 &\,=\frac{\del  }{\del \ln \PT^2} \frac{x}{x_*} G\left(\frac{x}{x_*}, \PT^2\right)\,+\,\frac{\alpha_s(\PT^2) N_c}{\pi}\int_{x_\star}^{1}\frac{\rmd\xi}{\xi}\,\frac{x}{\xi}
G\left(\frac{x}{\xi}, \PT^2\right),
\end{align} 
where in the last step we renamed $x_1\equiv x/x_*$ and we changed the integration  variable as $x'\equiv x/\xi$. The similarity with \eqn{DGLAP} becomes even more transparent if one denotes 
\beq
\pi \mcal{F}_g\left(\frac{x}{x_*}, \PT\right)\equiv \frac{\del  }{\del  \PT^2}\, \frac{x}{x_*} G\left(\frac{x}{x_*}, \PT^2\right)\,.\eeq
The last line of \eqn{DLA1} is independent of $x_*$ by construction. In our calculations, we go beyond the DLA in two different ways: we evaluate the unintegrated PDF $ \mcal{F}_g^{(0}(x/x_*,\PT)$ from the BK/JIMWLK evolution, and the gluon PDF $xG(x,\PT^2)$ from the DGLAP evolution. But these modifications have no impact at DLA: the respective contributions --- those that would be enhanced by powers of $ \alpha_s  \ln(1/x_*) \ln(\PT^2/\mu_0^2)$ --- are still compensating between the evolution of $ \mcal{F}_g^{(0}(x/x_*,\PT)$ and that of $xG(x,\PT^2)$. The residual $x_*$-dependence counts to order $\alpha_s\ln(1/x_*)\ll 1$, which is indeed negligible to the accuracy of interest --- a pure $\order{\alpha_s}$ effect.
 
 \subsection{TMD evolution with increasing $P_\perp$}
\label{sec:CSS}

The one-loop corrections in \eqn{FNLO} 
also describe an interesting evolution of the gluon TMD $\mcal{F}_g^{(1)} (x, K_{\perp}, P_\perp^2)$ with increasing $\PT$ at fixed $\KT$ that is encoded in the Sudakov logarithms. By taking a derivative w.r.t. $\ln\PT^2$, one finds
\begin{align}\label{CSS0}
\frac{\del \mcal{F}_g^{(1)} (x, K_{\perp}, P_\perp^2)}{\del \ln\PT^2}\,=\,
\frac{\alpha_s N_c }{2\pi}\left\{\frac{1}{\KT^2}\int^{\KT^2}_{\Lambda^2} {\dif\lt^2}\,
\mcal{F}_g^{(0)}(x, \ell_{\perp}) -\left(\ln\frac{\PT^2}{\KT^2}-2\beta_0\right)
\mcal{F}_g^{(0)}(x, K_{\perp}) \right\}.
 \end{align}
The generalisation of  this one-loop result to a genuine evolution equation is unambiguous if one requires this evolution to be Markovian (as generally the case with evolution equations in pQCD); that is, its r.h.s.~must be local in the evolution variable $\PT$. This condition implies 
\begin{align}
	\label{CSS}
 \frac{\del \mcal{F}_g(x, K_{\perp}, \PT^2)}{\del \ln\PT^2}=
& \, \frac{N_c}{2\pi}\left\{\frac{\alpha_s(K_\perp^2) }{K_\perp^2}
\int^{K_\perp^2}_{\Lambda^2}\rmd \ell_\perp^2\,\mcal{F}_g(x, \ell_{\perp}, \PT^2)
-\int_{K_\perp^2}^{P^2}\frac{\rmd \ell_\perp^2}{\ell_\perp^2}\alpha_s(\ell_\perp^2)\mcal{F}_g(x, K_{\perp}, \PT^2)\right\}\nonumber\\
&+\beta_0\frac{\alpha_s(P_\perp^2)N_c}{\pi}\mcal{F}_g(x, K_{\perp}, \PT^2)\,. \end{align}
In all the terms, $\alpha_s$ runs with the transverse momentum of the daughter gluons produced by the hard splitting. 

Eq.\,\eqref{CSS} describes the evolution of the gluon TMD when simultaneously increasing the transverse and the longitudinal resolution scales that are used to probe it (both controlled by $\PT$). The first two terms in the r.h.s. refer to the change in the longitudinal resolution: they describe the effects of  additional, soft, gluon emissions,  which become possible when increasing $\PT$. To see this, notice that the gluon emitted in the $s$-channel has a  longitudinal fraction  $1-\xi\ge \xi_0 =\KT/\PT$, where the lower limit decreases when increasing $\PT$. Hence, by increasing $\PT$, one allows for softer and softer emissions in the $s$-channel, which modify the distribution of the $t$-channel gluon in $\KT$, but not also in $x$
(which explains why Eq.\,\eqref{CSS} is local in $x$).
 

In the conformal limit $\beta_0\to 0$,
\eqn{CSS} reduces to a {\it rate equation}, with a ``gain term'' (corresponding to real emissions) and a ``loss term'' (the virtual ones). The number of gluons in the bin at $K_\perp$ can either
increase, via the process $\boldsymbol{\ell}_\perp\to (\bK, -\bK)$ with $\ell_\perp\ll \KT$, or decrease,
via the process $\bK\to (\bK', -\bK')$ with $\KT\ll \KT'\ll \PT$.  Via manipulations similar to those in  \eqn{rate0}, one can easily check that the total number of gluons, as defined in \eqn{PDF}, is not modified by this dynamics:
the ``gain'' and ``loss'' effects  mutually cancel:
\begin{align}\label{rate}
\int^{\PT^2}_{\Lambda^2} {\rmd K_\perp^2}\left\{\frac{\alpha_s(K_\perp^2) }{K_\perp^2}
\int^{K_\perp^2}_{\Lambda^2}\rmd \ell_\perp^2\,\mcal{F}_g(x, \ell_{\perp}, \PT^2)
-\int_{K_\perp^2}^{P^2}\frac{\rmd \ell_\perp^2}{\ell_\perp^2}\alpha_s(\ell_\perp^2)\mcal{F}_g(x, K_{\perp}, \PT^2)\right\}\,=\,0\,.\end{align}
Incidentally, this cancellation provides a non-trivial check of our previous assignments for the running couplings in \eqn{CSS} and for the scale ($\PT$) dependence of the TMD in the r.h.s. For instance, \eqn{rate} would not hold anymore if one was to replace $\mcal{F}_g(x, \ell_{\perp}, \PT^2)$ by $\mcal{F}_g(x, \ell_{\perp}, \KT^2)$ inside the integrand of the gain term. 

In real QCD where $\beta_0>0$, the last term in Eq.\,\eqref{CSS} acts as a driving force which increases the total number of gluons. This is consistent with the fact the coefficient of this last term is the anomalous dimension of the gluon distribution (the virtual contribution to the DGLAP splitting function); see also the discussion in  Sect.~\ref{sec:diag}.

The cancellation in \eqn{rate} is also necessary for the consistency between the definition Eq.~\eqref{PDF} of the gluon PDF and the DGLAP equation \eqn{DGLAP}. Indeed, by taking a derivative w.r.t. $ \ln\PT^2$ in \eqn{PDF}, one finds
\begin{align}\label{derG}
 \frac{\del xG  (x, \PT^2)}{\del \ln  \PT^2}&\,=\,\pi  \PT^2 \mcal{F}_g(x,  \PT,  \PT^2)+
 \pi \int^{ \PT^2}_{\Lambda^2} {\rmd K_\perp^2}\, \frac{\del \mcal{F}_g(x, K_{\perp},  \PT^2)}{\del \ln  \PT^2}
 \nn*[0.2cm] &\,=\,\pi  \PT^2 \mcal{F}_g(x,  \PT,  \PT^2)+\beta_0\frac{\alpha_s(P_\perp^2)N_c}{\pi}\,xG  (x, \PT^2),
 \end{align}
where in the second line we have used the fact that, after integrating \eqn{CSS} over $\KT^2$, 
it is only the $\beta_0$--piece that survives, cf. \eqn{rate}. The boundary value $\mcal{F}_g(x,  \PT^2,  \PT^2)$ is determined by \eqn{eq:Dreal} with $G^{(0)}\to G$, that is,
\begin{align}\label{Fboundary}
\mcal{F}_g(x,  \PT,  \PT^2)=\mcal{F}_g^{(0)}(x, P_{\perp}) + 
 \frac{\alpha_s }{2\pi^2}\frac{1}{P_\perp^2}
\int_{x_\star}^{1}\rmd\xi\,{P}^{(+)}_{gg}(\xi)\,\frac{x}{\xi}G\left(\frac{x}{\xi}, P_\perp^2\right)\,.
\end{align}
By combining Eqs.~\eqref{derG} and \eqref{Fboundary}, one recovers the DGLAP equation \eqn{DGLAP}, as anticipated.

\eqn{Fboundary} with $\PT^2\to \KT^2$ also serves as an initial condition (at $\PT^2= \KT^2$)
 for the rate equation Eq.~\eqref{CSS}. After using \eqn{derG} together with the one-loop running of the coupling, this boundary term can be alternatively written as
  \begin{align}
	\label{CSSinitcond}
	 \mcal{F}_g(x, \KT,  \KT^2) &\,=\, \frac{1}{\pi} \frac{\del x G(x,  \KT^2)}{\del  \KT^2}\,-\,
	 \frac{\beta_0 N_c}{\pi^2}\,\frac{\alpha_s( \KT^2) }{ \KT^2}\,xG(x, \KT^2)
	  \nonumber\\
  & \,=\, \frac{1}{\pi \alpha_s( \KT^2)} \,\frac{\del }{\del  \KT^2} \big[\alpha_s( \KT^2) \, x G(x,  \KT^2)\big]
  \,.\end{align}

\begin{figure}[t]
    \centering
    \includegraphics[width=0.7\textwidth]{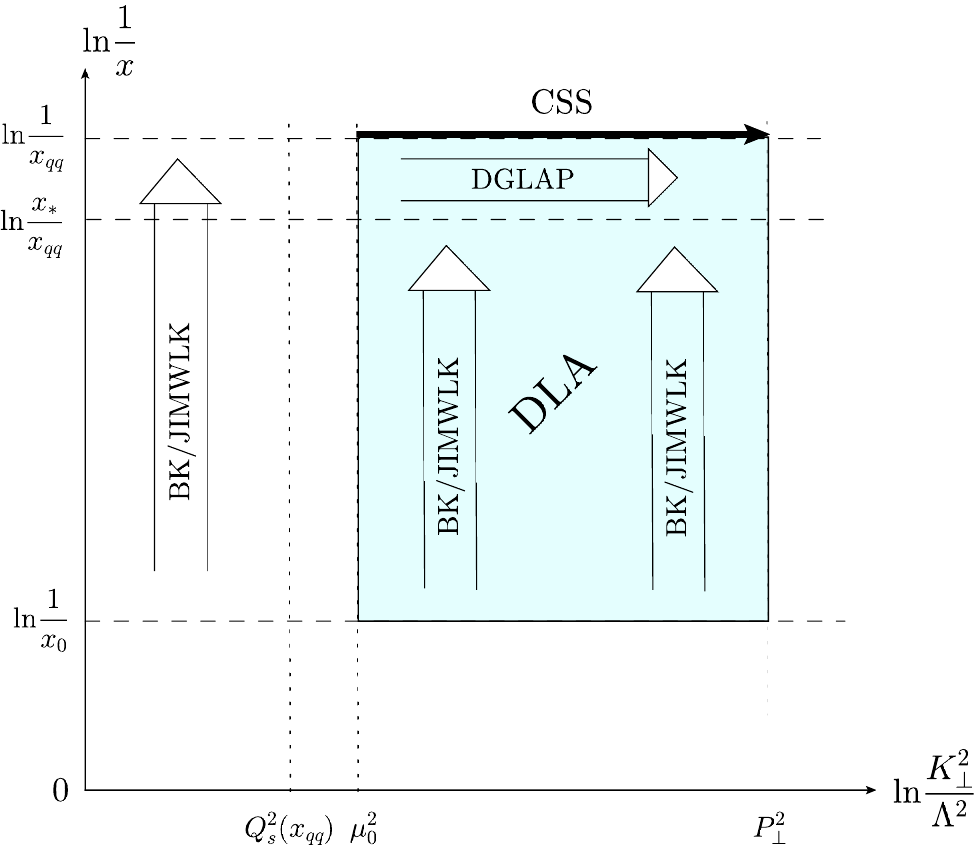}
     \caption{Graphical representation of the phase-space for  the three types of evolution which matter for dijet production at small $x_{q\bar q}$ and for relatively  large transverse momenta $\PT\gg\KT\gg Q_s(x_{q\bar q})$. The region which is peint in light blue is the domain where the double logarithmic approximation becomes valid. The source term in the DGLAP equation \eqn{DGLAP} is evaluated at $ x =x_{q\bar q}/x_*$ and for any $\KT$. The initial condition  \eqn{ICdglap} is formulated at $\KT =\mu_0$ and for $x$ within the range
    $x_{q\bar q}< x <x_{q\bar q}/x_*$. }
    \label{fig:cartoons}
\end{figure}

In the next couple of sections, we shall argue that \eqn{CSS} is a particular case of the  Collins-Soper-Sterman (CSS) equations for the gluon TMD~\cite{Collins:1981uk,Collins:1981uw,Collins:1984kg,Collins:2011zzd}. 
Before doing that, let us summarise here the recipe for combining together the three types of evolution 
--- BK/JIMWLK, DGLAP and CSS --- which emerge from the NLO corrections to the cross-section for hard dijet production as computed in the CGC effective theory. To {\it genuine} leading order in pQCD, that is, after resumming all the potentially large logarithms which formally appear at one-loop order, the dijet cross-section is given by the TMD-factorised expression in \eqn{sigma0} with the tree-level gluon WW TMD $\mcal{F}_g^{(0)}(x, \bK)$ replaced by the solution $ \mcal{F}_g(x, \bK,  \PT^2)$ to the following sequence of evolutions
(see Fig.~\ref{fig:cartoons} for a graphical illustration of the regions in phase-space that are covered by the various evolutions):

 \texttt{(i)} The BK/JIMWLK evolution of the gluon WW TMD with decreasing $x$ from $x=x_0\sim 10^{-2}$ (where one needs a non-perturbative initial condition, that can be taken from the  McLerran-Venugopalan model~\cite{McLerran:1993ni,McLerran:1993ka}) down to $x=x_{q\bar q}$. This step provides the source term $\mcal{F}_g^{(0)}(x_{q\bar q}/x_*, K_{\perp})$ which enters the r.h.s. of the DGLAP equation \eqn{DGLAP}, together with the initial condition \eqn{ICdglap} for any $x$. In practice, this  initial condition is needed only  for $x_{q\bar q}< x <x_{q\bar q}/x_*$, as clear from the r.h.s.~of \eqn{DGLAP}.
 
 \texttt{(ii)} The DGLAP evolution, cf.~\eqn{DGLAP},  with increasing $\KT^2$ from $\mu_0^2$, where  the initial condition  \eqn{ICdglap} is formulated, up to the  hard scale  $\PT^2$ (the dijet relative transverse momentum). This step provides the gluon PDF $xG(x, \KT^2)$ for  $x=x_{q\bar q}$ and any $\mu_0^2 < \KT^2 < \PT^2$. Via \eqn{CSSinitcond}, it also provides the initial condition (or boundary value) $ \mcal{F}_g(x, \KT,  \KT^2)$  for the CSS equation.

   \texttt{(iii)} The CSS evolution, cf.~\eqn{CSS},  with increasing $\PT^2$ from a generic value $\KT^2$
   (with $\KT^2 \ge \mu_0^2$) up to the physically interesting value $ \PT^2\gg \KT^2$. This step provides the final result $ \mcal{F}_g(x_{q\bar q}, \KT,  \PT^2)$ for the gluon TMD, to be used in the calculation of the dijet cross-section,  \eqn{sigma0}.

Clearly, all these steps are truly necessary in the situation where $x_{q\bar q}\ll x_0$ and  $ \PT^2\gg \KT^2
\gg Q_s^2(A,x_{q\bar q}) $.  One important example in that sense is dijet production in  ultra-peripheral Pb+Pb collisions (UPCs) at the LHC  \cite{ATLAS:2022cbd,CMS:2022lbi,ATLAS:2024mvt}. In this experimental set-up,
the measured jets have transverse momenta $P_\perp\ge 20$~GeV, a relatively large imbalance $\KT\gtrsim 5$~GeV, and their production probes the gluon distribution in the nuclear target at $x \lesssim 10^{-3}$.

Conversely, if  $x_{q\bar q}$ is not much smaller than $ x_0$, namely such that $\alpha_s\ln(x_0/x_{q\bar q})\ll 1$, then one can ignore the effects of the BK/JIMWLK evolution and, by the same token, also the rapidity divider $x_*$. In such a case, the source term in the DGLAP equation Eq.\,\eqref{DGLAP} is directly given by the MV model (so, it is independent of $x$), whereas the lower limit in the integral over $\xi$ should be replaced by $x/x_0$. That is, the DGLAP evolution occupies the whole rapidity phase-space, from $x_0$ down to $x=x_{q\bar q}$.


%
%

\subsection{TMD evolution in transverse coordinate representation}
\label{sub:TMD-cs}

The CSS equations for the evolution of the TMDs~\cite{Collins:1981uk,Collins:1981uw,Collins:1984kg,Collins:2011zzd} ---
and, more generally, the equations  describing successive parton branching in the presence of two large but widely separated momentum scales~\cite{Dokshitzer:1978hw,Dokshitzer:1978yd,Bassetto:1983mvz}
--- are generally written in transverse coordinate (``impact parameter'') space \cite{Parisi:1979se,Kodaira:1981nh,Ellis:1981sj,Collins:1981uk}.
That is, they describe the evolution of the Fourier transform $\tilde{\mcal{F}}_g(x,  \bbp,  \PT^2)$ of the gluon TMD, defined as
\beq\label{FTTMD}
\tF(x,  \bbp,  \PT^2)\,\equiv\int \frac{\rmd^2\bK}{(2\pi)^2}\,\rme^{i\bK\cdot\bbp}\, \mcal{F}_g(x, \bK, \PT^2)\,.
\eeq
There are several reasons which make this representation appealing.
In the context of the Parton Branching method for resumming Sudakov logarithms~\cite{Hautmann:2017xtx,Hautmann:2017fcj,BermudezMartinez:2023mrp}, the transverse coordinate formulation enables one to factorise the convolution introduced by transverse momentum conservation at each branching. Similarly, for processes in which more than one TMD are necessary to describe the initial state (like Drell-Yan where two quark TMDs are involved), using TMDs in coordinate space enables one factorise the momentum-space convolution~\cite{Boussarie:2023izj}. As we shall see, this formulation also allows to simplify the equation for the resolution scale dependence of the TMD and its non perturbative initial condition. Last, as explained after Eq.\,\eqref{gTMDV}, the Sudakov logarithms in the virtual term in the dense regime $K_\perp\sim Q_s$ naturally appear in coordinate space, therefore the coordinate space version of the CSS equation facilitates the matching between the fixed order NLO corrections and the resummed result in the saturation regime. Note that in the CSS formalism, this matching is also traditionally performed in coordinate space~\cite{Boussarie:2023izj}.

In what follows, we shall use \eqn{FTTMD} to transform our previous equation  \eqn{CSS} to impact parameter space. We first observe that the integration over $\KT$ in  \eqn{FTTMD} is effectively restricted to momenta $\KT < \PT$ by the support of the TMD, meaning that the transverse separation $\bt$ cannot be too small:  $\bt\gtrsim 1/P_\perp$.  Furthermore, since  $\rme^{i\bK\cdot\bbp}\simeq 1$ so long as $\KT\lesssim 1/\bt$,  it should be clear that 
the Fourier transform effectively integrates out (with unit weight) the low-momentum modes with $\KT< 1/\bt$. Since these modes are not explicitly measured anymore, their effect on the evolution of $\tilde{\mcal{F}}_g(x,  \bbp,  \PT^2)$  must cancel against  the respective virtual corrections. As a result, we expect the Fourier transform of \eqn{CSS} to involve only virtual contributions from the relatively hard modes with $\KT > 1/\bt$. This is indeed what we shall find.

Consider the two terms within the accolades of \eqn{CSS}; their two-dimensional Fourier transform reads (we temporarily suppress the argument $x$ of the TMD, since this plays no role for the subsequent manipulations)
\begin{align}
	& \int\frac{\rmd\KT^2}{4\pi}\,J_0(\bt\KT)\,\left[\frac{\alpha_s(\KT^2)}{K_\perp^2}
\int^{K_\perp^2}_{\Lambda^2}\rmd \ell_\perp^2\,\mcal{F}_g(\ell_{\perp}, \PT^2)
-\int_{K_\perp^2}^{\PT^2}\frac{\rmd \ell_\perp^2}{\ell_\perp^2}\, \alpha_s(\lt^2)
\mcal{F}_g(K_{\perp}, \PT^2)\right]
\end{align}
where we trivially performed the integral over the azimuthal angle between $\bK$ and $\bbp$.
As already mentioned, the integral over $\KT^2$ is effectively limited to $\KT^2 <\PT^2$. 
The lower cutoff $\Lambda$ (whenever needed) can  also be used to regularise the Fourier transform.
 After renaming  the integration variables in the first double integral (the ``gain'' term) and exchanging the order
 of integrations in the second double integral (the ``loss'' term),  we rewrite the above as
\begin{align}\label{CSSbt0}
	&\frac{1}{4\pi}\int_{\Lambda^2}^{\PT^2}\frac{\rmd \ell_\perp^2}{\ell_\perp^2}\, \alpha_s(\lt^2)
	\int_{\Lambda^2}^{\lt^2} \rmd\KT^2\,\mcal{F}_g(K_{\perp}, \PT^2)\left[
	J_0(\bt\lt)-J_0(\bt\KT)\right].
\end{align}
The two Bessel functions are both equal to one (and hence cancel each other) when $\lt \ll 1/\bt$, whereas for larger values $\lt\gg 1/\bt$, we can keep only the second Bessel function (the first one is rapidly oscillating, so it gives a negligible contribution to the integral over $\lt^2$); to summarise,
\beq
\Theta(\lt-\KT)\left[	J_0(\bt\lt)-J_0(\bt\KT)\right]\,\simeq\,- \Theta(\lt-\KT)\Theta(\lt-1/\bt) J_0(\bt\KT)\,.\eeq
After using this, \eqn{CSSbt0} simplifies to
\begin{align}\label{CSSrealvirt}
	-\int_{1/\bt^2}^{\PT^2}\frac{\rmd \ell_\perp^2}{\ell_\perp^2}\, \alpha_s(\lt^2)
	\,\tF(\bt, \PT^2).\end{align}
where we have recognised the Fourier transform of the gluon TMD. Notice that the upper limit $\lt^2$ on the integral over $\KT^2$ did not prevent us from reconstructing the Fourier tranform according to  \eqn{FTTMD}. Indeed since $\lt$ is restricted  to relatively large values $\lt\gg 1/\bt$, whereas $\KT$ is limited to $\KT\lesssim 1/\bt$ by the Bessel function 
$J_0(\bt\KT)$, the upper cutoff $\lt^2$ plays no role in  the integral over $\KT^2$.

After also adding the piece proportional to $\beta_0$ in the r.h.s. of \eqn{CSS} (whose Fourier transform is trivial), we obtain the 
 impact parameter version of our equation:
 \begin{align}\label{CSSimpact}
 \frac{\del{\tF}(x, \bt, \PT^2)}{\del\ln \PT^2}=&\frac{N_c}{\pi}
 \left[-\frac{1}{2}\int_{c^2_0/\bt^2}^{\PT^2} \frac{\rmd\lt^2}{\ell_\perp^2}\, \alpha_s(\lt^2)
+\beta_0\alpha_s(\PT^2) \right]\tF(x, \bt, \PT^2),\end{align}
with $c_0$ a number of order one.  The value $c_0=2\,e^{-\gamma_E}\simeq 1.123$ is often chosen in the literature, as it minimises the error introduced when going from \eqn{CSSbt0} to \eqn{CSSrealvirt} \cite{Ellis:1981sj}. In addition, when solving Eq.\,\eqref{CSSimpact} at the  order $\alpha_s$ (one iteration only), this choice for $c_0$ enables one to match the exact NLO expression for the Sudakov logarithms in the saturation regime, following the discussion after Eq.\,\eqref{gTMDV}. As anticipated, there is no explicit real (``gain'') term in \eqn{CSSimpact}:
the real piece that was originally present in its momentum version  \eqn{CSS}  has disappeared after the Fourier transform, while leaving behind the essential condition $\lt>1/\bt$ in the above integral.

\eqn{CSSimpact} can be easily solved for an arbitrary initial condition $\tilde{\mcal{F}}_0(x, \bt)$ (the value of 
${\tF}(x, \bt, \PT^2)$ for $\PT= c_0/\bt$). The solution reads
  \begin{align}\label{CSSimpactsol}
{\tF}(x, \bt, \PT^2)\,=\,\tilde{\mcal{F}}_0(x, \bt)\exp
\left\{- \frac{ N_c}{\pi}\int_{\mu_b^2}^{\PT^2} \frac{\rmd\lt^2}{\ell_\perp^2}\,
 \alpha_s(\lt^2)\,\left[\frac{1}{2}\ln\frac{\PT^2}{\lt^2}-\beta_0\right]\right\},
  \end{align}
where we have introduced the notation $\mu_b^2\equiv c^2_0/\bt^2$  for the lower limit.
  

\subsection{Relation to the CSS equation}
\label{sec:diag}

As previously explained, an important effect of the quantum corrections to the cross-section is to render
the gluon TMD dependent upon the hard scale $P_\perp$ (the relative momentum of the $q\bar q$ dijet), which effectively acts as a resolution scale for measuring gluons in the target wavefunction. Remarkably, this unique scale controls both the transverse and the longitudinal resolutions used to discriminate the gluon emissions: indeed, the  $P_\perp$--dependence of the NLO TMD enters via the phase-space boundaries on both the gluon transverse momentum $\kg$ and its longitudinal momentum fraction $z_g$ (or $1-\xi$ in the target picture), cf. Eqs.~\eqref{Fqq}-\eqref{FSukreal}-\eqref{gTMDreal}. 

This situation should be compared to the more familiar description of the (quark and gluon) TMDs, which generally involves {\it two} distinct resolution scales: a transverse factorisation scale $\mu$ introduced by the ultraviolet renormalisation (which also controls the running of the coupling) and a longitudinal scale $\zeta$ related to the subtraction of the rapidity divergence. Both scales appear since, in the standard approach, the one-loop (and higher) corrections to the TMD are computed from its operator definition\footnote{Strictly speaking, at operator level one needs to consider both the operator defining the bare TMD and that for the soft function~\cite{Collins:2011zzd,Boussarie:2023izj}.}, without reference to a specific scattering process. (See the handbooks~\cite{Collins:2011zzd,Boussarie:2023izj} for pedagogical examples.) In such a general set-up, the two types of divergences must be separately regulated and subtracted, which naturally leads to two independent resolution scales. These scales are {\it a priori} arbitrary, so the physical cross-sections should be independent of them. This condition leads to two evolution equations, describing the TMD evolution with varying the resolution scales, often referred to as the Collins-Soper-Sterman (CSS) equations~\cite{Collins:1981uk,Collins:1981uw,Collins:1984kg,Collins:2011zzd}. (The same equations have also been obtained with the SCET formalism~\cite{Becher:2010tm,Becher:2012yn,Echevarria:2011epo,Echevarria:2014rua,Chiu:2012ir,Ebert:2019tvc}.) 

The precise definition of $\zeta$ (and hence the precise form of respective evolution equation) depends upon the scheme used for regulating the rapidity divergence (see
\cite{Collins:2017oxh,Ebert:2022cku,Boussarie:2023izj} for comparisons between different approaches). For definiteness, we shall use Collins-11 scheme \cite{Collins:2011zzd}, in which the gluon TMD in position space
 is written as
$\tilde F_g(x, \bt, \mu, \zeta)$ where both $\mu$ and  $\zeta$ have the dimension of mass\footnote{Note a small difference w.r.t. the conventions in  \cite{Collins:2011zzd}: we use the notation $\zeta^2$ for the quantity denoted there as $\zeta$.}.
The evolution equations for the gluon TMD w.r.t. the two resolution scales take the general structure  \cite{Collins:2011zzd}:
  \begin{align}\label{CSS11}
  \frac{\del \ln \tilde F_g(x, \bt, \mu, \zeta)}{\del\ln\mu^2}&=
  \frac{1}{2}\gamma_G\big(\alpha_s(\mu),\zeta/\mu\big)=
   \frac{1}{2}\left\{\gamma_G\big(\alpha_s(\mu),1\big)-\Gamma_G\big(\alpha_s(\mu)\big)\,\ln\frac{\zeta^2}{\mu^2}\right\}\,,\\*[0.2cm]
    \frac{\del \ln \tilde F_g(x, \bt, \mu, \zeta)}{\del\ln\zeta^2}&= \frac{1}{2}\tilde K\big(\bt, \mu\big)
    =- \frac{1}{2}\int_{\mu_b^2}^{\mu^2} \frac{\rmd\lt^2}{\ell_\perp^2}\,\Gamma_G\big(\alpha_s(\lt)\big)
    +\frac{1}{2}\tilde K\big(\bt, \mu_b\big).
  \end{align}
Here, $\gamma_G$ is the anomalous dimension for the UV renormalisation of the gluon TMD, 
$\Gamma_G$ is the cusp anomalous dimension for the adjoint representation, and $\tilde K$ is the Collins-Soper kernel (see~\cite{Collins:2011zzd,Boussarie:2023izj} for details). We recall that $\mu_b^2=c^2_0/\bt^2$ with  $c_0=2\,e^{-2\gamma_E}$. To LO in $\alpha_s$ one has (see e.g. \cite{Echevarria:2015uaa})
\beq
\Gamma_G^{(0)}\big(\alpha_s(\mu)\big)\,=\,\frac{\alpha_s(\mu)N_c}{\pi}\,,\qquad
\gamma_G^{(0)}\big(\alpha_s(\mu),1\big)\,=\,2\beta_0\frac{\alpha_s(\mu)N_c}{\pi}\,,\qquad
\tilde K^{(0)}\big(\bt, \mu_b\big)\,=\,0,
\eeq
hence the CSS equations Eq.~\eqref{CSS11} simplify to
  \begin{align}\label{CSS11LO}
  \frac{\del \ln \tilde F_g(x, \bt, \mu, \zeta)}{\del\ln\mu^2}&\simeq\,\frac{\alpha_s(\mu)N_c}{\pi}\,
 \left\{\beta_0- \frac{1}{2}\ln\frac{\zeta^2}{\mu^2}\right\}\,,\\*[0.2cm]
    \frac{\del \ln \tilde F_g(x, \bt, \mu, \zeta)}{\del\ln\zeta^2}&\simeq - \frac{N_c}{2\pi}
    \int_{\mu_b^2}^{\mu^2} \frac{\rmd\lt^2}{\ell_\perp^2}\,\alpha_s(\lt)\,.  \end{align}
Let us introduce  the ``diagonal'' limit of the above TMD in which the two resolution scales are identified with each other and they are both equal to some external scale $\PT$ (as we shall shortly argue, the gluon TMD that we have studied throughout this paper is a distribution of this type):
\beq\label{TMDdiag}
 \tilde F_g^{\rm D}(x, \bt, \PT)\,\equiv\,\tilde F_g(x, \bt, \mu=\PT, \zeta=\PT).\eeq
 The evolution of this new distribution w.r.t. $\PT$ is clearly given by 
  \begin{align}\label{CSS11D}
  \frac{\del \ln \tilde F_g^{\rm D}(x, \bt, \PT)}{\del\ln\PT^2}&=
  \frac{1}{2}\left\{\gamma_G\big(\alpha_s(\PT),1\big)+\tilde K\big(\bt, \PT\big)\right\}
  \\*[0.2cm] &\,  \simeq
  \,\beta_0\frac{\alpha_s(\PT)N_c}{\pi}- \frac{N_c}{2\pi}
    \int_{\mu_b^2}^{\PT^2} \frac{\rmd\lt^2}{\ell_\perp^2}\,\alpha_s(\lt)
    \,,  \end{align}
where the approximate equality holds to LO in $\alpha_s$. Clearly, the final equation is identical to that satisfied by our distribution ${\tF}(x, \bt, \PT^2)$, cf.~\eqn{CSSimpact}. Hence, the gluon TMD emerging from our NLO calculation of the dijet cross-section obeys the ``diagonal'' version of the CSS equations. It can therefore be identified with $ \tilde F_g^{\rm D}(x, \bt, \PT)$ provided the initial condition at $\PT=\mu_b$ is correspondingly chosen:
\beq\label{ident}
{\tF}(x, \bt, \PT^2)\,=\,\tilde F_g^{\rm D}(x, \bt, \PT).\eeq

Alternatively, if one independently evolves in $\mu$ and $\zeta$, by solving the general CSS equations Eq.~\eqref{CSS11}, and then of identifies the resolution scales both in the initial condition, 
$\mu_0=\zeta_0\equiv \mu_b$, and at the end of the evolution, $\mu=\zeta\equiv \PT$, one obtains exactly the same result as evolving $\tilde F_g^{\rm D}(x, \bt, \PT)$  according to \eqn{CSS11D}, from an initial value $\mu_b$ up to the final value $\PT$ of interest. Indeed, the general solution to the CSS equations with two resolution scales reads (see also Eq.~(4.17) in \cite{Boussarie:2023izj})
 \begin{align}\label{CSSsol1}
\tilde F_g(x, \bt, \mu, \zeta)&\,=\,\tilde F_0(x, \bt, \mu_0, \zeta_0)\,\exp\left\{
\frac{1}{2}\int_{\mu_0^2}^{\mu^2} \frac{\rmd\lt^2}{\ell_\perp^2}\left[
\gamma_G\big(\alpha_s(\lt),\zeta/\lt\big)\right]\right\}\,\exp\left\{\frac{1}{2}
\tilde K\big(\bt, \mu_0\big)\ln\frac{\zeta^2}{\zeta_0^2}\right\}.
\end{align} 
In the ``diagonal'' limit $\mu=\zeta\equiv \PT$ and $\mu_0=\zeta_0\equiv \mu_b$, this becomes
\begin{align}\label{CSSsolD}
\ln\frac{\tilde F_g^D(x, \bt, \PT)}{\tilde F_g^D(x, \bt, \mu_b)}
\,=\ \frac{1}{2}\int_{\mu_b^2}^{\PT^2} \frac{\rmd\lt^2}{\ell_\perp^2}\left[
\gamma_G\big(\alpha_s(\lt),1\big)-\Gamma_G\big(\alpha_s(\lt)\big)\,\ln\frac{\PT^2}{\lt^2}\right]
+ \frac{1}{2}\tilde K\big(\bt, \mu_b\big) \ln\frac{\PT^2}{\mu_b^2}.
\end{align}
This result should be compared to the solution to  \eqn{CSS11D}, easily found as
\begin{align}\label{CSSDsol}
\ln\frac{\tilde F_g^D(x, \bt, \PT)}{\tilde F_g^D(x, \bt, \mu_b)}
&\,=\ \frac{1}{2}\int_{\mu_b^2}^{\PT^2} \frac{\rmd\lt^2}{\ell_\perp^2}\left[
\gamma_G\big(\alpha_s(\lt),1\big)+\tilde K\big(\bt, \lt\big)\right]
\nn*[0.2cm]&
\,=\ \frac{1}{2}\int_{\mu_b^2}^{\PT^2} \frac{\rmd\lt^2}{\ell_\perp^2}\left[
\gamma_G\big(\alpha_s(\lt),1\big) -\int_{\mu_b^2}^{\lt^2} \frac{\rmd\mu^{\prime 2}}{\mu^{\prime 2}}
\,\Gamma_G\big(\alpha_s(\mu')\big)\right]+
\frac{1}{2}\tilde K\big(\bt, \mu_b\big) \ln\frac{\PT^2}{\mu_b^2}.
\end{align}
At this point, it is easy to verify that the double integral over $\lt$ and $\mu'$ in the last equation indeed gives the same result as the piece involving $\Gamma_G$ in \eqn{CSSsolD}. This establishes the equivalence between the two solutions.

Let us now explain why this identification is indeed natural for the problem at hand.
We first observe that unlike in the traditional, ``bottom-up'', approach,  to TMD factorisation \cite{Collins:2011zzd,Boussarie:2023izj}, in which the loop corrections are separately computed for the partonic TMDs (starting with their operator definitions) and for the hard factor, and the results are subsequently combined to demonstrate the cancellation of the arbitrary factorisation scales $\mu$ and $\zeta$ in the cross-section (see e.g.~Refs.~\cite{Ji:2004wu,Ji:2005nu,Sun:2011iw} and, for the specific case of inclusive dijet in DIS, see \cite{delCastillo:2020omr}), in our ``top-down'' approach, we compute directly  the cross-section for which we demonstrate TMD factorisation via controlled approximations. In this process, there is no need to introduce arbitrary factorisation scales: the one-loop corrections to the dijet cross-section show no UV, nor collinear divergences, and all the potential divergences are cut off by physical scales. Hence, the gluon TMD which emerges as a building block of the TMD factorisation is only sensitive to such physical scales.

Notice that also in the traditional ``bottom-up'' approach, the factorisation scales like $\mu$ and $\zeta$ are eventually replaced by physical scales in the result for the cross-section. For instance, it is customary to chose $\mu^2=\PT^2$ (the largest transverse momentum scale in the problem), in order to avoid large transverse logarithms in the hard factor. 
In fact, all the calculations of Sudakov factors in the CGC formalism at small $x$ that we are aware of --- from the pioneering papers \cite{Mueller:2012uf,Mueller:2013wwa} to the recent NLO calculations  \cite{Taels:2022tza,Caucal:2023nci,Caucal:2023fsf,Altinoluk:2024vgg} --- involve a single hard scale (the Higgs mass,  the photon virtuality, the dihadron relative transverse momentum ....) which controls both the upper limit on the transverse momentum integration (``UV cutoff'') and the rapidity cutoff.

While the identification of the UV factorisation scale $\mu$ with dijet relative  transverse momentum $\PT$ may indeed look natural, one may wonder why the rapidity factorisation scale (the Collins-Soper scale $\zeta$) can be identified with $\PT$ as well. To understand that, let us first recall the definition of this scale in the standard TMD approach: in the Collins-11 scheme~\cite{Collins:2011zzd}, $\zeta$ is defined in terms of
the upper cutoff $y_n$ on the rapidity of the emitted gluon, as follows
 (see e.g. Eq.~(10.127) in Ref. [11] that we adapt to the kinematics at hand)
\begin{equation}\label{defzeta}
\zeta^2 \equiv  P_\perp^2 \,e^{2(y-y_n)},
\end{equation}
where $P_\perp$ and $y$ are the transverse momentum and the rapidity of the emitter, assumed to be on-shell. In our case (and in the projectile picture), the emitter is a quark or an antiquark, with rapidity (say for the quark)
\beq
y_1= \frac{1}{2}\ln\frac{k_1^+}{k_1^-}=\ln\frac{\sqrt{2} z_1 q^+}{k_{1\perp}}\simeq 
\ln\frac{\sqrt{2} q^+}{P_{\perp}}\,,\eeq
where we have used $z_1\sim \order{1}$ and $k_{1\perp}\simeq P_\perp$. (A similar estimate holds for the antiquark.) In our dijet problem, the upper cutoff  $y_n$ on the rapidity $y_g$ of the emitted gluon is determined by the maximal valued allowed for $z_g$, which in turn is determined by the condition that the gluon be emitted at large angles, outside the jet cone. This requires  $z_g \lesssim z_n \equiv k_{g\perp}/P_\perp$, which implies an upper limit on $y_g$:
\begin{equation}
y_g =\frac{1}{2}\ln\frac{k_g^+}{k_g^-}=\ln\frac{\sqrt{2} z_g q^+}{k_{g\perp}}\ \lesssim \ y_n=
\ln\frac{\sqrt{2}z_n q^+}{k_{g\perp}}\,=\,
\ln\frac{\sqrt{2} q^+}{P_{\perp}}\,\simeq y_1\,.
\end{equation}
We thus see that $y_n\simeq y_1$, which together with \eqn{defzeta} implies  $\zeta=P_\perp$, as anticipated.

%

Besides the fact that it naturally emerges from our calculation, this diagonal version of the TMD has another virtue that started being appreciated only recently in the TMD literature: it allows one to construct the  PDF by simply integrating the TMD over $\KT$ up to the resolution scale $\PT$. This is of course the formula \eqn{PDF} that we have used on extenso in our previous analysis (and which in particular allowed us to construct the DGLAP equation from the one-loop corrections to the gluon TMD). Yet, with the general definition of the TMD which involves two factorisation scales $\mu$ and $\zeta$, the existence of a simple relation like \eqn{PDF} is highly non-trivial (see e.g.~\cite{Ji:2004wu} for a discussion of the difficulties inherent in such an identification). Let us define the ``cumulative gluon TMD''
\beq\label{cum}
xg(x, \PT^2, \mu, \zeta)\,\equiv\,\pi\int_{\Lambda^2}^{\PT^2} \rmd\KT^2\,F_g(x, \KT, \mu, \zeta),\eeq
where the integrand involves the gluon TMD in momentum space (in the Collins-11 scheme). In Ref.~\cite{Ebert:2022cku}, this quantity has been numerically computed (as a function of $\PT$ for different values of $\mu/\PT$ and $\zeta/\PT$) and the result was compared with numerical solutions to the DGLAP equation for $xG(x,\PT^2)$. A fastidious analysis involving CSS solutions in position space and their Fourier transform led to the conclusion that the cumulative gluon TMD coincides (within numerical errors) with the standard PDF only for the special choice $\mu=\zeta=\PT$ (see Fig.~8 in \cite{Ebert:2022cku}).
Very recently, analytical arguments which support this numerical observation have been presented in \cite{delRio:2024vvq}. These arguments are quite similar to our earlier discussion in Sect.~\ref{sec:CSS}: they demonstrate (via pQCD calculations up to three-loop order) that the CSS and the DGLAP evolutions are consistent with each other provided the PDF and the TMD are related   via \eqn{cum} with  $\mu=\zeta=\PT$.

To conclude, our ``top-down'' construction of the TMD starting with the dijet cross-section  in the colour dipole picture not only provides an ``economical'' version of the gluon TMD, which involves a single resolution scale and obeys a simpler version of the CSS equations, but also allows for the simple relation Eq.\,\eqref{PDF} between the TMD and the respective PDF. Such a relation is of course natural at tree-level, but our analysis demonstrates that it is maintained by the quantum, DGLAP and CSS, evolutions, provided one uses the ``diagonal'' version of the TMD which naturally emerges from our analysis.

\subsection{Solving the CSS equation in momentum space}
\label{sec:CSSsol}

In this section, we will construct an analytic solution for the CSS equation in momentum space,  \eqn{CSS},
corresponding to a particularly simple initial condition
which mimics the  perturbative tail of the gluon TMD at large transverse momenta $\KT\gg Q_s(x)$.  The main interest of this solution is to illustrate the effects of the CSS resummation in a very explicit way. We shall also present for comparison two solutions obtained by working in impact parameter space and numerically computing the inverse Fourier transform $\bbp\to \bK$.

The analytic solution will be constructed via iterations and that purpose, it is preferable to use the integral version of the CSS equation,  which reads
\begin{align}
	\label{gTMDintfin}
	\mcal{F}_g(x, K_{\perp}, Q^2) = \mcal{F}_g(x, K_{\perp},K_{\perp}^2) 
	+\Delta \mcal{F}_g^{\rm Sud}(x, K_{\perp}, Q^2),
 \end{align}
The boundary term $\mcal{F}_g(x, K_{\perp},K_{\perp}^2)$ can be read off \eqn{CSSinitcond} and the Sudakov contribution $\Delta \mcal{F}_g^{\rm Sud}(x, K_{\perp}, Q^2)$ is obtained by integrating the r.h.s. of \eqn{CSS} over $\PT^2$ from $\KT^2$ up to $Q^2$:
 \begin{align}
	\label{gSUDint}\hspace{-0.8cm}
 \Delta \mcal{F}_g^{\rm Sud}(x, K_{\perp}, Q^2)=
& \, \frac{N_c}{2\pi}\,\frac{\alpha_s(K_\perp^2) }{K_\perp^2}\int_{\KT^2}^{Q^2}\frac{\rmd\PT^2}{\PT^2}
\int^{K_\perp^2}_{\Lambda^2}\rmd \ell_\perp^2\,\mcal{F}_g(x, \ell_{\perp}, \PT^2) - \frac{N_c}{2\pi}
\int_{K_\perp^2}^{Q^2}\frac{\rmd \ell_\perp^2}{\ell_\perp^2}\alpha_s(\ell_\perp^2)
\int_{\lt^2}^{Q^2}\frac{\rmd\PT^2}{\PT^2}\,\mcal{F}_g(x, K_{\perp}, \PT^2)
 \nonumber \\
& +\frac{\beta_0N_c}{\pi}\int_{\KT^2}^{Q^2}\frac{\rmd\PT^2}{\PT^2}\,\alpha_s(P_\perp^2)
\mcal{F}_g(x, K_{\perp}, \PT^2)\,.\end{align}

To simplify our analysis, we ignore the DGLAP evolution of the boundary term. So, the latter reduces to the initial condition, $\mcal{F}_g(x, K_{\perp},K_{\perp}^2)=\mcal{F}_0(x, K_{\perp})$, 
for which we shall take the particularly simple form
\beq\label{F0}
\mcal{F}_0(x, K_{\perp}) =\frac{1}{\KT^2}\,,\qquad  xG_0(x, Q^2) =\pi\ln\frac{Q^2}{\Lambda^2}\,.\eeq
Also, to simplify the writing, we present the details of the calculation for the case of a  fixed coupling $\abarz \equiv  {\alpha_s N_c}/{2\pi}$ and for $\beta_0=0$. Under these assumptions, \eqn{gTMDintfin} reduces to
  \begin{align}
	\label{gTMDint0}
	\mcal{F}_g(x, K_{\perp}, Q^2) =  \mcal{F}_0(x, K_{\perp}^2) +
	 \frac{\abarz}{K_\perp^2}\int_{\KT^2}^{Q^2}\frac{\rmd\PT^2}{\PT^2}
\int^{K_\perp^2}_{\Lambda^2}\rmd \ell_\perp^2\,\mcal{F}_g(x, \ell_{\perp}, \PT^2)
-\abarz \int_{K_\perp^2}^{Q^2}\,\frac{\rmd\PT^2}{\PT^2}\,
\ln\frac{\PT^2}{K_{\perp}^2}\,
\mcal{F}_g(x, K_{\perp}, P_\perp^2).\end{align}
As discussed in relation with  \eqn{rate}, the simplified evolution described by \eqn{gTMDint0} preserves the total number of gluons:
\begin{align}\label{intCSSsol}
\int_{\Lambda^2}^{Q^2} \rmd K_{\perp}\,\mcal{F}_g(x, K_{\perp}^2, Q^2)=
\ln\frac{Q^2}{\Lambda^2}=\int_{\Lambda^2}^{Q^2} \rmd K_{\perp}^2\,\mcal{F}_0(x, K_{\perp})\,.\end{align}
This can be used  to rewrite the real term in \eqn{gTMDint0} as : 
\beq
 \frac{\abarz}{K_\perp^2}\int_{\KT^2}^{Q^2}\frac{\rmd\PT^2}{\PT^2}
\int^{K_\perp^2}_{\Lambda^2}\rmd \ell_\perp^2\,\mcal{F}_g(x, \ell_{\perp}, \PT^2)
= \frac{\abarz}{K_\perp^2}\int_{\KT^2}^{Q^2}\frac{\rmd\PT^2}{\PT^2}
\left(\ln\frac{\PT^2}{\Lambda^2} - \int_{K_\perp^2}^{P_\perp^2}\rmd \ell_\perp^2\,\mcal{F}_g(x, \ell_{\perp}, \PT^2)\right)
\eeq
So, the integral equation becomes (after some simple manipulations)
 \begin{align}
	\label{gTMDint1}
	\mcal{F}_g(x, K_{\perp}^, Q^2)&\, = \mcal{F}_0(x, K_{\perp}^2) +
	\frac{\abarz}{ K_\perp^2} \left\{\ln\frac{Q^2}{K_{\perp}^2}\ln\frac{K_\perp^2}{\Lambda^2}
+\frac{1}{2} \ln^2\frac{Q^2}{K_{\perp}^2}\right\}
	 \\
&
	- \frac{\abarz}{K_\perp^2}\int_{\KT^2}^{Q^2}\frac{\rmd\PT^2}{\PT^2}
\int_{K_\perp^2}^{P_\perp^2}\rmd \ell_\perp^2\,\mcal{F}_g(x, \ell_{\perp}, \PT^2)
-\abarz \int_{K_\perp^2}^{Q^2}\,\frac{\rmd\PT^2}{\PT^2}\,
\ln\frac{\PT^2}{K_{\perp}^2}\,
\mcal{F}_g(x, K_{\perp}, P_\perp^2).  \nonumber\end{align}

We define iterations such that the $n$th order contribution $\mcal{F}_g^{(n)}$ is of order $\abarz^n$. Hence
$\mcal{F}_g^{(0)}=\mcal{F}_0$, while the first two iterations are given by
\begin{align}
K_{\perp}^2\,\mcal{F}_g^{(1)}(x, K_{\perp}, Q^2)\, &=
\abarz \ln\frac{\KT^2}{\Lambda^2}\, \ln\frac{Q^2}{K_{\perp}^2}
-\frac{\abarz}{2}\ln^2\frac{Q^2}{K_{\perp}^2}\,,\\
K_{\perp}^2\,\mcal{F}_g^{(2)}(x, K_{\perp}, Q^2)\, &=
- \frac{\abarz^2}{2} \ln\frac{\KT^2}{\Lambda^2}\, \ln^3\frac{Q^2}{K_{\perp}^2}
+ \frac{\abarz^2}{8} \, \ln^4\frac{Q^2}{K_{\perp}^2}\,
\,.
\end{align}
It is then easy to infer the form of the general term of order $n$:
\begin{align}\label{iter01}
K_{\perp}^2\,\mcal{F}_g^{(n)}(x, K_{\perp}, Q^2)&\, = (-1)^n\left\{
- \frac{\abarz^n}{2^{n-1}(n-1)!} \ln\frac{\KT^2}{\Lambda^2}\, \ln^{2n-1}\frac{Q^2}{K_{\perp}^2}
+ \frac{\abarz^n}{2^n n!} \, \ln^{2n}\frac{Q^2}{K_{\perp}^2}\right\}\,.
\end{align}
The two series generated in this way are both recognised as expansions of the same exponential, namely
\begin{align}\label{CSSsol}
\,\mcal{F}_g(x, K_{\perp}, Q^2)&\, =\,\frac{1}{K_{\perp}^2}
\left(1+\abarz \ln\frac{\KT^2}{\Lambda^2}\, \ln\frac{Q^2}{K_{\perp}^2}\right)
\,\exp\left\{-\frac{\abarz}{2}\ln^2\frac{Q^2}{K_{\perp}^2}\right\}.
\end{align}
As a check, one can use this solution to verify the ``sum-rule'' \eqn{intCSSsol} 
\begin{align}\label{intCSSsol1}
\int_{\Lambda^2}^{Q^2} \rmd K_{\perp}^2\,\mcal{F}_g(x, K_{\perp}, Q^2)=\int_0^{t_0}\rmd t \,\rme^{-\abarz (t_0-t)^2/2}
\Big[1+\abarz(t_0-t)t\Big]=t_0\equiv \ln\frac{Q^2}{\Lambda^2}.
\end{align}
where we set $t=\ln(\KT^2/\Lambda^2)$ and the final result is easily obtained after integrating by parts.


\eqn{CSSsol} has a very suggestive structure: it is the same as the result of one iteration of the real term in \eqn{gTMDint0} times an exponential which resums the virtual corrections to all orders. This suggests the following structure for the general solution, corresponding to an arbitrary initial condition $\mcal{F}_0(x, K_{\perp}^2)$:
\begin{align}\label{CSSsolgen}
\,\mcal{F}_g(x, K_{\perp}, Q^2)&\, =\,\left[\mcal{F}_0(x, K_{\perp}^2) + \frac{\abarz}{\pi K_{\perp}^2}
\ln\frac{Q^2}{K_{\perp}^2}\,xG_0(x,K_\perp^2)\right]\,\exp\left\{-\frac{\abarz}{2}\ln^2\frac{Q^2}{K_{\perp}^2}\right\}.
\end{align}
Let us first check that this expression preserves indeed the gluon number, as it should (cf.~\eqn{intCSSsol})
\begin{align}\label{intgensol}
\int_{\Lambda^2}^{Q^2} \rmd K_{\perp}^2\,\mcal{F}_g(x, K_{\perp}, Q^2)&\,=
\int_{\Lambda^2}^{Q^2} \rmd K_{\perp}^2\,
\left[\mcal{F}_0(x, K_{\perp}) + \frac{\abarz}{K_{\perp}^2}
\ln\frac{Q^2}{K_{\perp}^2}\int_{\Lambda^2}^{K_{\perp}^2} \rmd \ell_{\perp}^2\,
\mcal{F}_0(x, \ell_{\perp})\right]\,\exp\left\{-\frac{\abarz}{2}\ln^2\frac{Q^2}{K_{\perp}^2}\right\}
\nn*[0.2cm]
&\,=\int_{\Lambda^2}^{Q^2} \rmd K_{\perp}^2\,\frac{\del}{\del K_\perp^2}\left[
\left(\int_{\Lambda^2}^{K_{\perp}^2} \rmd \ell_{\perp}^2\,
\mcal{F}_0(x, \ell_{\perp})\right)\,\exp\left\{-\frac{\abarz}{2}\ln^2\frac{Q^2}{K_{\perp}^2}\right\}\right]
\nn*[0.2cm]
&\,=\int_{\Lambda^2}^{Q^2} \rmd \ell_{\perp}^2\,\mcal{F}_0(x, \ell_{\perp})\,,
\end{align}
which is the expected result. Via similar manipulations, it is furthermore easy to check that \eqn{CSSsolgen}
 is indeed a solution to the CSS equation Eq.~\eqref{CSS} for $\beta_0=0$ and a fixed coupling.

\begin{figure}[t] 
\centerline{\includegraphics[width=0.48\columnwidth,page=1]{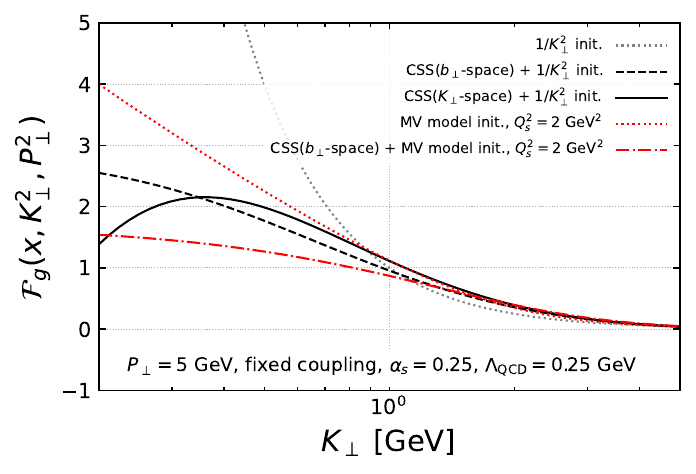}\quad
\includegraphics[width=0.49\columnwidth,page=2]{plot-CSS-btvskt-space-v2.pdf}}
\caption{\small The analytic solution in \eqn{CSSsol}, which is obtained by working in transverse momentum space and corresponds to the perturbative initial condition $ \mcal{F}_0(\KT)=1/\KT^2$,  is compared to  numerical solutions obtained in the $b_\perp$-representation for two different initial conditions: 
a perturbative initial condition with non-perturbative cutoff at large $b_\perp$, cf.~Eqs.~\eqref{FTT0}-\eqref{CSSimpact0FT}, and the MV model initial condition \eqn{FMV}, which includes saturation effects but no non-perturbative prescription.
The results are shown for two values of the resolution scale: $\PT= 5$~GeV (left)  and $\PT= 20$~GeV (right).}
\label{fig:CSS}
\end{figure}

\eqn{CSSsol} makes it clear that the effects of the Sudakov resummation are particularly important at relatively small $K_\perp\ll Q$: unlike the initial condition $ \mcal{F}_0(x, K_{\perp}) $ which is divergent as $\KT\to 0$, the CSS solution \eqn{CSSsol} vanishes in that limit.  On the other hand, when $\KT$ approaches $Q$ from the below, the Sudakov double logarithm becomes less important, so we recover the tail in  $1/K_\perp^2$. This discussion shows that the function in \eqn{CSSsol} must have a maximum at some intermediate value $\Lambda < \KT < Q$. A rough estimate for the position of this maximum can be obtained  by using a simplified version of  \eqn{CSSsol}, where we ignore the (slowly-varying) logarithmic term in the middle factor: this yields  $ K_{\rm max}^2 \sim Q^2 e^{-1/\abarz}\ll Q^2$. This qualitative behaviour is confirmed by the curves shown in full line in Fig.~\ref{fig:CSS}, which exhibit the solution  \eqn{CSSsol} as a function of $\KT$ for $\alpha_s=0.2$ and the two values for $\PT$: 5 GeV and 20~GeV. A maximum is indeed visible, but it occurs at very low transverse momenta, of order $\Lambda$, hence the nearby behaviour cannot be really trusted. Yet, the strong reduction in $\mcal{F}_g$ due to the CSS evolution is already visible at much larger momenta $\KT\gg K_{\rm max}$, so this is a robust (and expected) consequence 
of this resummation.


It is furthermore interesting to compare the (analytic) solution in momentum space,  \eqn{CSSsol},
with the solution corresponding to the same physical initial condition, but obtained by working in impact parameter space. We recall that the $b_\perp$-space solution is given by  \eqn{CSSimpactsol}, which under present assumptions reduces to
 \begin{align}\label{CSSimpact0}
 \tF(x, \bt, Q^2)\,=\,\tilde{\mathcal{F}}_0(x, \bt)\,\exp\left\{-\frac{\abarz}{2}\ln^2 \frac{Q^2 \bt^2}{c_0^2}\right\},
  \end{align}
with  $c_0=2\,e^{-\gamma_E}\simeq 1.123$. We first choose the
initial condition $\tilde{\mathcal{F}}_0(x, \bt)$ as the Fourier transform of \eqn{F0}:
\beq\label{FTT0}
\tilde{\mathcal{F}}_0(x, \bt)\,\equiv\int \frac{\rmd^2\bK}{(2\pi)^2}\,\rme^{i\bK\cdot\bbp}\, \frac{1}{\KT^2}\,=\,
\frac{1}{2\pi}\,\ln\frac{c_0}{\bt\Lambda}\,\to\,\frac{1}{2\pi}\,K_0(\Lambda \bt).
\eeq
This is strictly defined only for $\bt\ll 1/\Lambda$, but its extension to large, non-perturbative, values $\bt\gtrsim 1/\Lambda$ is needed when computing the inverse Fourier transform (from $\bt$ to $\KT$). To that purpose, we replace the logarithm with the  modified Bessel function $K_0(\Lambda \bt)$, which exponentially vanishes when  $\bt\gtrsim 1/\Lambda$. The corresponding gluon TMD is obtained by numerically computing the following integral,
 \begin{align}\label{CSSimpact0FT}
 \mcal{F}_g(x, K_{\perp}, Q^2)
 \,=\int \frac{\rmd^2\bbp}{2\pi}\,\rme^{i\bK\cdot\bb}\,K_0(\Lambda \bt)\,
\exp \left\{-\frac{\abar}{2}\ln^2  \frac{Q^2 \bt^2}{c_0^2}\right\}\,.
  \end{align}
with results shown by the dashed curves in Fig.~\ref{fig:CSS}. These curves show no maximum, in agreement with the fact that the behaviour at very low momenta $\KT\lesssim\Lambda$ has been regulated by the exponential cutoff (via the $K_0$ modified Bessel function) at large values of $\bt$.

A particularly useful feature of the  impact parameter representation is the fact that it allows one to  easily include the effects of gluon saturation (generally computed as multiple scattering in the eikonal approximation). As a simple illustration, let us use the initial condition for the WW gluon TMD provided by the McLerran-Venugopalan (MV) model~\cite{McLerran:1993ni,McLerran:1993ka}, that is~\cite{Dominguez:2011wm,Lappi:2017skr,Dumitru:2016jku,Boussarie:2021ybe}
\beq
\label{FMV}
\tilde{\mcal{F}}_{\rm MV}(x, \bt)\,=\,\frac{1}{2\pi Q_0^2}\frac{1-e^{-\Gamma(\bt)}}{\Gamma(\bt)}\left[\frac{\rmd^2}{\rmd b_\perp^2}+\frac{1}{b_\perp}\frac{\rmd}{\rmd b_\perp}\right]\Gamma(b_\perp)\,,\qquad\,\Gamma(\bt)=\frac{\bt^2 Q_0^2}{4}
\ln\left(\frac{1}{\bt\Lambda_{\rm MV}}+e\right),\eeq
where the normalisation 
and the non-perturbative scale $\Lambda_{\rm MV}=\Lambda e^{-1+\gamma_E}/2$ are chosen in such a way to match the logarithmic behaviour in \eqn{FTT0} at small $\bt\ll 1/Q_s$. The quantity $T_{\rm MV}(\bt)\equiv  1-e^{-\Gamma(\bt)}$ is the elastic scattering amplitude of a gluon-gluon dipole as computed in the MV model.
We recall that the saturation momentum $Q_s$ in this context is defined as the value $Q_s=2/\bt$ at which the exponential in Eq.\,\eqref{FMV} becomes of order one:
\beq
Q_s^2=Q_0^2 \,\ln\left(\frac{Q_s}{2\Lambda_{\rm MV}}+e\right)\,.\eeq
Notice that in this case there is no need for any non-perturbative prescription at large $\bt\gg 1/Q_s$: the respective behaviour is already cut off by saturation. After inserting this initial condition in \eqn{CSSimpact0} and numerically performing the inverse Fourier transform, we obtain the results shown with dashed-dotted (red) lines in  Fig.~\ref{fig:CSS}, for the case where $Q_s^2=2$~GeV$^2$. As expected, the softening of the spectrum (i.e.~the deviation from a rapid increase $\propto 1/\KT^2$ at low $\KT$) is now manifest already for larger momenta $\KT\sim Q_s\gg\Lambda$.

\section{Conclusions and outlook}
\label{sec:concl}

To summarise, we have demonstrated that the CGC approach to  $\gamma^* A$ interactions at small $x$ not only is consistent with TMD factorisation for back-to-back dijet production, but also 
generates the collinear quantum evolutions expected for the gluon TMD in the regime where both the dijet relative momentum $P_\perp$ and the dijet imbalance $\KT$ are much larger than the nuclear saturation momentum $Q_s(A,x)$. 
In this context, we have clarified the interplay between the DGLAP/CSS evolutions of the gluon PDF/TMD and the high-energy, BK/JIMWLK, evolution of the Weiszäcker-William unintegrated gluon distribution. The various evolutions  refer to complementary regions in phase-space, as illustrated in Fig.~\ref{fig:cartoons}, and do not overlap with each other  to the accuracy of interest.

Our analysis focused on dijet (as opposed to dihadron) final states and this choice has important consequences for the rapidity phase-space available to gluon emissions at NLO: the gluons must be soft enough to be emitted at large angles. In turn, this condition fixed the coefficient of the Sudakov double logarithm to be the same as the cusp anomalous dimension for gluons, which eventually allowed us to obtain the expected version of the CSS equation for the gluon TMD in the target. Hence, our results confirm (in this particular context) the important, but rarely verified~\cite{Sun:2014gfa,Caucal:2024vbv}, expectation that the Sudakov effects for jet production can be resummed by solving the CSS equations for initial-state (i.e. prior to the collision) parton TMDs.

Another interesting aspect of our analysis is that it provides a   ``top-down'' approach to TMD factorisation at small $x$: unlike the standard (``bottom-up'') approach at moderate $x$, where TMD factorisation is postulated at tree-level and then shown to be consistent with the higher-order corrections, in our approach this factorisation emerges from the CGC calculations, first at tree-level, then at one-loop order, via controlled approximations. In this approach, there are no arbitrary resolution scales, rather they are all fixed by the physical context. For the dijet problem at hand, it turns out that the same, hard, momentum scale --- the dijet relative momentum $\PT$ --- plays both the role of a ultraviolet cutoff and that of the Collins-Soper scale. This identification between the two resolution scales (transverse and longitudinal) has the benefit to preserve at one-loop level the relation  between the gluon PDF $xG(x, \PT^2)$ and the gluon TMD $\mcal{F}_g(x, K_{\perp}, \PT^2)$ that is familiar at tree-level: the former can be computed by integrating the latter over  $\KT$ up to $\PT$. Via this relation (plus an appropriate boundary condition at $\PT=\KT$), the DGLAP equation and the diagonal version of the CSS equation are consistent with each other.

To our knowledge, it is for the first time such a unified calculation scheme is shown to emerge from a first principle calculation at small $x$. 
Similar results can be obtained for other processes, like $\gamma$-jet production in $pA$ collisions, semi-inclusive and diffractive jet production in DIS. Such processes include additional types of partonic TMDs for the target: the dipolar gluon TMD~\cite{Dominguez:2010xd,Dominguez:2011wm}, the sea quark TMD~\cite{Marquet:2009ca,Xiao:2017yya,Caucal:2024vbv}, and the diffractive quark and gluon TMDs~\cite{Iancu:2021rup,Hatta:2022lzj,Iancu:2022lcw,Hauksson:2024bvv}.
They will be discussed in separate publications, together with applications to the phenomenology. The strategy and the methodology that we proposed here  will hopefully allow for systematic studies of particle production with large transverse momenta in dilute-dense collisions at small $x$.

\begin{acknowledgments}
\noindent{\bf Acknowledgements.} We are grateful to Al Mueller and Feng Yuan for inspiring discussions. We would like to thank Yoshitaka Hatta, Sigtryggur Hauksson and Farid Salazar for useful remarks. 
We thank the France-Berkeley-Fund from University of California at Berkeley for support and the Nuclear Theory Group at the Lawrence Berkeley National Laboratory for its hospitality during the completion of this project. E.I. thanks the EIC Theory Institute at the Brookhaven National Laboratory for hospitality and support during the final stages of this work.
Figures were created with JaxoDraw ~\cite{Binosi:2003yf}.
 \end{acknowledgments}

\let\oldaddcontentsline\addcontentsline
\renewcommand{\addcontentsline}[3]{}
\bibliographystyle{apsrev4-1}
\bibliography{refs}
\let\addcontentsline\oldaddcontentsline

\end{document}